\documentclass[11pt,a4paper]{article}

\usepackage{subfigure,jcappub,amsfonts,amsmath,amssymb,bm,graphicx}

\title{Gravitational waves generation in turbulent hypermagnetic fields before the electroweak phase transition}

\author{Maxim Dvornikov}
\emailAdd{maxdvo@izmiran.ru}
\affiliation{Pushkov Institute of Terrestrial Magnetism, Ionosphere
and Radiowave Propagation (IZMIRAN),
108840 Moscow, Russia}

\abstract{
We study the production of relic gravitational waves (GWs) in
turbulent hypermagnetic fields (HMFs) in the symmetric phase of the
early universe before the electroweak phase transition (EWPT). The
noise of HMFs is modeled by the analog of the magnetic hydrodynamics
turbulence. The evolution of HMFs is driven by the analogs of the chiral
magnetic effect and the Adler anomalies in the presence of the nonzero
asymmetries of leptons and Higgs bosons. We track the evolution of
the energy density of GWs from $10\,\text{TeV}$ down to EWPT and
analyze its dependence on the parameters of the system. We also discuss
the possibility to observe the predicted GW background by the current
GW detectors.}
\begin{document}

\maketitle

\section{Introduction}

The observation of gravitational waves (GWs) by the LIGO-Virgo collaborations~\cite{Abb16}
was one of the direct confirmations of the validity of the General
Relativity. Since then, the records of the numerous sources of GWs
have been collected~\cite{Abb21}. There are vast plans~\cite{Bai21}
to build new GW detectors in addition to the currently operating ones.

Besides the GW signal emitted by certain astrophysical objects, there
should be stochastic GWs filling all the universe~\cite{Chr19}.
The detection of stochastic GWs is quite challenging because of the
rather low frequencies of the signal. Some of the detection methods
are reviewed in ref.~\cite{RomCor17}.

There are numerous sources of the stochastic GW background. They can
be of the astrophysical origin~\cite{Reg11} or stem from the early
universe~\cite{CapFig18}. Even for the cosmological GW background, one
can point out on multiple sources. Relic GWs can be produced
in the inflation epoch, during various phase transitions, by cosmic
strings etc. Many of the mechanisms are reviewed in ref.~\cite{CapFig18}.

We are interested in the generation of primordial GWs by turbulent
magnetic fields (see, e.g., ref.~\cite{KosMacKah02}). However, instead
of dealing with a magnetic field, we shall treat the production of GWs
before the electroweak phase transition (EWPT). In this case, instead
of a photon $A^{\mu}$, one has the massless hypercharge field $Y^{\mu}=\sin\theta_{\mathrm{W}}Z^{\mu}+\cos\theta_{\mathrm{W}}A^{\mu}$,
where $Z^{\mu}$ is the $Z$-boson and $\theta_{\mathrm{W}}$ is the
Weinberg angle. Thus, we replace the electromagnetic field tensor
with $F_{\mu\nu}\to(\mathbf{E}_{\mathrm{Y}},\mathbf{B}_{\mathrm{Y}})$,
where $\mathbf{E}_{\mathrm{Y}}$ and $\mathbf{B}_{\mathrm{Y}}$ are
hyperelectric and hypermagnetic fields (HMF).

All leptons can be treated as massless particles before EWPT. Moreover,
one can have nonzero lepton asymmetries $n_{f}\neq n_{\bar{f}}$ in
this epoch. Thus, we can apply the analog of the chiral magnetic effect
(the CME)~\cite{Fuk08} for the description of the HMF evolution.
The CME consists in the excitation of an electric current of massless
particles, forming plasma with a nonzero chiral imbalance, along an
external magnetic field. The magnetic field in this situation becomes
unstable. There is a backreaction from the helical magnetic field
to the particle asymmetry resulting from the Adler anomaly (see, e.g.,
ref.~\cite{PesSch95}). It leads to the decay of a helical magnetic
field and the production a nonzero asymmetry of massless fermions.
For the first time, this scenario was used for the leptogenesis in
the early universe in ref.~\cite{JoySha97}.

The mechanism, which involves the CME and the Adler anomaly, was used
mainly in connection to the lepto- and baryogenesis in (hyper-)magnetic
fields in the early universe (see, e.g., refs.~\cite{DvoSem13,KamLon16}).
However, we can treat evolving HMFs as a source of primordial GWs.
The production of GWs driven by unstable magnetic fields accounting
for the chiral anomalies was considered recently in refs.~\cite{Bra21,Pan21}.

In ref.~\cite{DvoSem21}, we studied the influence of the HMF turbulence
on the lepto- and baryogenesis in the early universe before EWPT.
The noise of HMFs was modeled by the hypermagnetic hydrodynamics [(H)MHD]
turbulence. In this approach, the plasma motion in the Navier-Stokes
equation was driven mainly by the Lorentz force~\cite{Sig02}. To
calculate the baryon asymmetry of the universe (BAU) we were interested
in the evolution of the lepton asymmetries. Now, if we apply the model
in ref.~\cite{DvoSem21} to describe the GWs production by the HMF
turbulence, we will deal with the evolution of the spectra of the
HMF energy density and helicity.

This work is organized in the following way. In section~\ref{sec:PRODGW},
we rederive the energy density of GWs and its spectrum in turbulent
HMFs in the symmetric phase of the early universe before EWPT. We
summarize the dynamics of HMFs, as well as the lepton and Higgs boson
asymmetries in section~\ref{sec:HMFEVOL}. The results of the numerical
simulations are present in section~\ref{sec:RES}. The possibility to
observe predicted relic GWs with the modern GWs detectors is considered
in section~\ref{subsec:OBS}. We conclude in section~\ref{sec:CONCL}.
The main kinetic equations are rewritten in the form convenient for
numerical simulations in appendix~\ref{sec:NEWVAR}.

\section{Production of GWs\label{sec:PRODGW}}

We consider the tensor perturbation $h_{\mu\nu}$ of the background
Friedmann--Robertson--Walker metric $\bar{g}_{\mu\nu}=\text{diag}(1,-a^{2},-a^{2},-a^{2})$,
where $a=a(t)$ is the scale factor. The interval reads now
\begin{equation}
  \mathrm{d}s^{2}=g_{\mu\nu}\mathrm{d}x^{\mu}\mathrm{d}x^{\nu},\quad g_{\mu\nu}=\bar{g}_{\mu\nu}+h_{\mu\nu},
\end{equation}
where we use the physical coordinates $x^{\mu}=(t,\mathbf{x})$. We
can express the spatial part of $h_{\mu\nu}$ as $h_{ij}=a^{2}D_{ij}$,
where the matrix $D_{ij}$ is supposed to vary harmonically $D_{ij}(\mathbf{x},t)\sim e^{\mathrm{i}\mathbf{kx}}D_{ij}(\mathbf{k},t)$.
Here $\mathbf{k}=a\mathbf{k}_{\mathrm{phys}}$ is the conformal momentum.
Using the transverse-traceless gauge, in which $D_{ii}=0$ and $k_{i}D_{ij}=0$,
one gets that $D_{ij}$ obeys the equation~\cite{Wei20},
\begin{equation}\label{eq:Deq}
  -a^{2}\ddot{D}_{ij}-3a\dot{a}\dot{D}_{ij}+\nabla^{2}D_{ij}=16\pi G\pi_{ij}^{(\mathrm{T})},
\end{equation}
where $\pi_{ij}^{(\mathrm{T})}=P_{ikjl}T_{kl}$, $T_{kl}$ is the
energy-momentum tensor, $P_{ikjl}=P_{ik}P_{jl}-\tfrac{1}{2}P_{ij}P_{kl}$,
$P_{ij}=\delta_{ij}-\hat{k}_{i}\hat{k}_{j}$, $\hat{\mathbf{k}}=\mathbf{k}/k$
is the unit vector, $G=M_{\mathrm{Pl}}^{-2}$ is the Newton constant,
and $M_{\mathrm{Pl}}=1.2\times10^{19}\,\text{GeV}$ is the Planck
mass. We can see that $k_{i}\pi_{ij}^{(\mathrm{T})}=0$ and $\pi_{ii}^{(\mathrm{T})}=0$.

We suppose that metric perturbations are caused by the noise of HMF
in the symmetric phase before EWPT. In this case, $T_{\mu\nu}$ reads
\begin{equation}
  T_{\mu\nu}=\frac{1}{4}\bar{g}_{\mu\nu}F_{\alpha\beta}^{(\mathrm{Y})}F_{\mathrm{Y}}^{\alpha\beta}-
  F_{\mu\alpha}^{(\mathrm{Y})}F_{\nu\beta}^{(\mathrm{Y})}\bar{g}^{\alpha\beta},
\end{equation}
where $F_{\mu\nu}^{(\mathrm{Y})}=\partial_{\mu}Y_{\nu}-\partial_{\nu}Y_{\mu}$
is the tensor of the hyperfield potential $Y^{\mu}$. Considering
length scales greater than the Debye radius, we get that the hyperelectric
field is screened effectively. The energy-momentum tensor takes the
form,
\begin{equation}\label{eq:Tij}
  T_{ij}=-\frac{1}{a^{2}}\left(B_{\mathrm{Y}i}^{(c)}B_{\mathrm{Y}j}^{(c)}-\frac{1}{2}\delta_{ij}B_{\mathrm{Y}}^{(c)2}\right),
\end{equation}
where $\mathbf{B}_{\mathrm{Y}}^{(c)}=a^{2}\mathbf{B}_{\mathrm{Y}}$
is the conformal HMF.

Using eq.~(\ref{eq:Tij}), we rewrite eq.~(\ref{eq:Deq}) in the
form,
\begin{equation}\label{eq:Dprime}
  D_{ij}^{\prime\prime}+2HD_{ij}^{\prime}+k^{2}D_{ij}=f_{ij},
  \quad
  f_{ij}=\frac{16\pi G}{a^{2}}P_{ikjl}
  \left(
    B_{\mathrm{Y}k}^{(c)}B_{\mathrm{Y}l}^{(c)}-\frac{1}{2}\delta_{kl}B_{\mathrm{Y}}^{(c)2}
  \right),
\end{equation}
where $H=a'/a$ is the Hubble constant and the prime means the derivative
with respect to the conformal time $\eta$ defined by $\mathrm{d}\eta=\mathrm{d}t/a$.
Since we are at the radiation dominated universe, we can choose $a(t)=\sqrt{t/t_{\text{Univ}}}$,
where $t_{\text{Univ}}=1.4\times10^{10}\,\text{yr}$ is the age of
the Universe.\footnote{Strictly speaking, the choice of the scale factor is not exact since there is a matter dominated universe presently. The current scale factor reads $a_{\mathrm{matt}}\propto t^{2/3}$ which modifies the normalization constant in $a_{\mathrm{rad}}\propto\sqrt{t}$.} It corresponds to the scale factor $a_{\text{now}}=a(t=t_{\text{Univ}})=1$
nowadays. The conformal time is chosen such that $\eta(t)=2\sqrt{t_{\text{Univ}}}(\sqrt{t}-\sqrt{t_{*}})$,
where $t_{*}$ is the initial time when perturbations start growing.
Note that $\eta(t_{*})=0$. Equation~(\ref{eq:Dprime}), rewritten
using new variables, takes the form,
\begin{equation}\label{eq:Dg}
  D_{ij}''+\frac{2}{\eta+\eta_{0}}D_{ij}'+k^{2}D_{ij}=f_{ij}(\mathbf{k},\eta),
\end{equation}
where $\eta_{0}=2\sqrt{t_{\text{Univ}}t_{*}}$. We supply eq.~(\ref{eq:Dg})
with the initial condition $D_{ij}(\mathbf{k},0)=D'_{ij}(\mathbf{k},0)=0$.
Note that $\eta+\eta_{0}=2t_{\text{Univ}}a(\eta)$. The general solution, satisfying
the chosen initial condition, reads
\begin{equation}\label{eq:Dsol}
  D_{ij}(\eta)=\frac{1}{ka(\eta)}
  \int_{0}^{\eta}\frac{\sin(k[\eta-\eta_{1}])}{a(\eta_{1})}f_{ij}^{(c)}(\mathbf{k},\eta_{1})\mathrm{d}\eta_{1},
\end{equation}
where $f_{ij}^{(c)}=a^{2}f_{ij}$.

Equation~(\ref{eq:Dsol}) is used to compute the energy density of
GWs $\rho_{\mathrm{GW}}$ which is the time component $\rho_{\mathrm{GW}}=t_{00}$
of the effective energy-momentum tensor of GW~\cite{LanLif88},
$t_{\mu\nu}=\frac{1}{32\pi G}\left\langle \partial_{\mu}h_{\alpha\beta}\partial_{\nu}h^{\alpha\beta}\right\rangle $.
The conformal energy density reads
\begin{equation}
  \rho_{\mathrm{GW}}^{(c)}=a^{4}\rho_{\mathrm{GW}}=
  \frac{a^{2}}{32\pi G}
  \left\langle
    \frac{\mathrm{d}}{\mathrm{d}\eta}(a^{2}D_{ij})\cdot\frac{\mathrm{d}}{\mathrm{d}\eta}(a^{-2}D_{ij})
  \right\rangle.
\end{equation}
Assuming that the time scale of a perturbation is shorter than the
expansion rate of the universe, we can write down the spectral density
of GWs as 
\begin{align}\label{eq:rhockern}
  \rho_{\mathrm{GW}}^{(c)}(\mathbf{k},\eta)=&
  \delta^{3}(\mathbf{k}-\mathbf{k}')\frac{1}{8\pi G}\int_{0}^{\eta}
  \int_{0}^{\eta}\frac{\mathrm{d}\eta_{1}\mathrm{d}\eta_{2}}{a(\eta_{1})a(\eta_{2})}
  \notag
  \\
  &\times
  \left\langle
    \cos(k[\eta-\eta_{1}])\cos(k[\eta-\eta_{2}])f_{ij}^{(c)}(\mathbf{k},\eta_{1})f_{ij}^{(c)*}(\mathbf{k}',\eta_{2})
  \right\rangle.
\end{align}
We change the variables in the double integral in eq.~(\ref{eq:rhockern})
to $\eta_{1}=\xi+\xi'$ and $\eta_{2}=\xi$. We also separate the
averaging between cosines and $f_{ij}^{(c)}$. The integral in eq.~(\ref{eq:rhockern})
transforms to
\begin{multline}\label{eq:xixi'}
  \int_{0}^{\eta}d\xi\int_{-\xi}^{\eta-\xi}\frac{\mathrm{d}\xi'}{a(\xi+\xi')a(\xi)}
  \left\langle
    \cos(k[\eta-\xi-\xi'])\cos(k[\eta-\xi])
  \right\rangle 
  \\
  \times
  \left\langle
    f_{ij}^{(c)}(\mathbf{k},\xi+\xi')f_{ij}^{(c)*}(\mathbf{k}',\xi)
  \right\rangle. 
\end{multline}
Unlike ref.~\cite{KosMacKah02}, where the generation of GWs was
studied during a relatively short time after a phase transition, we
neglect the variable $\xi'$ in the inner integrand in eq.~(\ref{eq:xixi'})
since we study the production of GWs in a long time interval. The
average of the cosines factor is $\left\langle \cos^{2}(k[\eta-\xi])\right\rangle =1/2$.
Eventually, the spectral density of GWs takes the form,
\begin{equation}\label{eq:rhocfin}
  \rho_{\mathrm{GW}}^{(c)}(\mathbf{k},\eta)=
  \delta^{3}(\mathbf{k}-\mathbf{k}')\frac{\eta}{16\pi G}
  \int_{0}^{\eta}\frac{\mathrm{d}\xi}{a^{2}(\xi)}
  \left\langle
    f_{ij}^{(c)}(\mathbf{k},\xi)f_{ij}^{(c)*}(\mathbf{k}',\xi)
  \right\rangle.
\end{equation}
Note that $\rho_{\mathrm{GW}}^{(c)}(\mathbf{k},\eta)$ is proportional
to $\eta$ as in ref.~\cite{KosMacKah02}.

The function $f_{ij}^{(c)}$ is the binary combination of the hypermagnetic
components. To compute $f_{ij}^{(c)}$, we should, first, determine
$\left\langle B_{\mathrm{Y}i}^{(c)}(\mathbf{k},\eta)B_{\mathrm{Y}j}^{(c)*}(\mathbf{k}',\eta)\right\rangle $.
We take that~\cite{Cam07,DvoSem17}
\begin{equation}\label{eq:correl}
  \left\langle
    B_{\mathrm{Y}i}^{(c)}(\mathbf{k},\eta)B_{\mathrm{Y}j}^{(c)*}(\mathbf{k}',\eta)
  \right\rangle =
  (2\pi)^{3}\delta^{3}(\mathbf{k}-\mathbf{k}')\frac{1}{2}
  \left[
    (\delta_{ij}-\hat{k}_{i}\hat{k}_{j})S_{\mathrm{Y}}(k,\eta)-\mathrm{i}\varepsilon_{ijn}\hat{k}_{n}A_{\mathrm{Y}}(k,\eta)
  \right],
\end{equation}
where $S_{\mathrm{Y}}(k,\eta)=(2\pi)^{2}\rho_{\mathrm{Y}}^{(c)}(k,\eta)/k^{2}$,
$A_{\mathrm{Y}}(k,\xi)=2\pi{}^{2}h_{\mathrm{Y}}^{(c)}(k,\eta)/k$,
$\rho_{\mathrm{Y}}^{(c)}(k,\eta)$ is the spectrum of the hypermagnetic
energy density, and $h_{\mathrm{Y}}^{(c)}(k,\eta)$ is the spectrum
of hypermagnetic helicity density. Both $\rho_{\mathrm{Y}}^{(c)}(k,\eta)$
and $h_{\mathrm{Y}}^{(c)}(k,\eta)$ are defined in conformal variables.
The total hypermagnetic energy and the hypermagnetic helicity are
given by $\rho_{\mathrm{Y}}^{(c)}(\eta)\equiv B_{\mathrm{Y}}^{(c)2}/2=\smallint\mathrm{d}k\rho_{\mathrm{Y}}^{(c)}(k,\eta)$
and $h_{\mathrm{Y}}^{(c)}(\eta)\equiv\tfrac{1}{V}\smallint\mathrm{d}^{3}x(\mathbf{Y}^{(c)}\cdot\mathbf{B}_{\mathrm{Y}}^{(c)})=\smallint\mathrm{d}kh_{\mathrm{Y}}^{(c)}(k,\eta)$.

Using eq.~(\ref{eq:correl}), after lengthy but straightforward calculations,
we get that
\begin{align}\label{eq:ffcorr}
  \left\langle
    f_{ij}^{(c)}(\mathbf{k},\xi)f_{ij}^{(c)*}(\mathbf{k}',\xi)
  \right\rangle = &
  \delta(\mathbf{k}-\mathbf{k}')\frac{32G^{2}}{\pi}
  \notag
  \\
  & \times
  \int\frac{\mathrm{d}^{3}q}{(2\pi)^{3}}
  \big[
    (1+\cos^{2}\alpha)(1+\cos^{2}\beta)S_{\mathrm{Y}}(|\mathbf{q}|,\xi)S_{\mathrm{Y}}(|\mathbf{k}-\mathbf{q}|,\xi)
    \notag
    \\
    &+
    4\cos\alpha\cos\beta A_{\mathrm{Y}}(|\mathbf{q}|,\xi)A_{\mathrm{Y}}(|\mathbf{k}-\mathbf{q}|,\xi)
  \big],
\end{align}
where $\cos\alpha=(\hat{k}\cdot\hat{q})$ and $\cos\beta=(\hat{k}\cdot\widehat{k-q})$.
Finally, with help of eqs.~(\ref{eq:rhockern}) and~(\ref{eq:ffcorr}),
we obtain the expression for the spectral density of GWs in the form,
\begin{align}\label{eq:rhoGWisotr}
  \rho_{\mathrm{GW}}^{(c)}(k,\eta)= &
  \frac{t_{\text{Univ}}^{2}G}{4k^{3}\pi^{2}}\eta\int_{0}^{\eta}\frac{\mathrm{d}\xi}{(\eta_{0}+\xi)^{2}}
  \int_{0}^{\infty}\frac{\mathrm{d}q}{q^{3}}\int_{|k-q|}^{k+q}\frac{\mathrm{d}p}{p^{3}}
  \nonumber \\
  & \times
  \big\{
    [4k^{2}q^{2}+(k^{2}+q^{2}-p^{2})^{2}][4k^{2}p^{2}+(k^{2}-q^{2}+p^{2})^{2}]
    \rho_{\mathrm{Y}}^{(c)}(q,\xi)\rho_{\mathrm{Y}}^{(c)}(p,\xi)
    \nonumber
    \\
    & +
    4k^{2}q^{2}p^{2}(k^{2}+q^{2}-p^{2})(k^{2}-q^{2}+p^{2})
    h_{\mathrm{Y}}^{(c)}(q,\xi)h_{\mathrm{Y}}^{(c)}(p,\xi)
  \big\}.
\end{align}
The total conformal energy density can be obtained using eq.~(\ref{eq:ffcorr}) as
\begin{equation}\label{eq:densGW}
  \rho_{\mathrm{GW}}^{(c)}(\eta)=\int_{0}^{\infty}\rho_{\mathrm{GW}}^{(c)}(k,\eta)\mathrm{d}k.
\end{equation}
We use the assumption of the spectra isotropy to derive eqs.~(\ref{eq:rhoGWisotr})
and~(\ref{eq:densGW}).

To produce relic GWs, we use random HMFs with the spectrum in the
range $k_{\mathrm{min}}<k<k_{\mathrm{max}}$ (see section~\ref{sec:HMFEVOL}
below). In this case, basing on eq.~(\ref{eq:rhoGWisotr}), one gets
that spectrum of GWs produced is in the interval $0<k<2k_{\mathrm{max}}$.

\section{HMFs evolution\label{sec:HMFEVOL}}

According to eq.~(\ref{eq:rhoGWisotr}), the GWs generation is owing
to the presence of the nonzero $\rho_{\mathrm{Y}}^{(c)}(k,\xi)$ and
$h_{\mathrm{Y}}^{(c)}(k,\xi)$. We have to choose the appropriate
model for the evolution of HMFs.

In the symmetric phase of the universe evolution before EWPT, the
behavior of HMFs can be driven by the analog of the CME in the presence
of nonzero lepton asymmetries. This model was studied in details in
ref.~\cite{DvoSem21} in connection to the BAU problem. The full
set of the kinetic equations reads,
\begin{align}\label{eq:HMFsys}
  \frac{\partial\tilde{\mathcal{E}}_{{\rm B_{\mathrm{Y}}}}}{\partial\tilde{\eta}} & =
  -2\tilde{k}^{2}\eta_{\mathrm{eff}}\tilde{\mathcal{E}}_{{\rm B_{\mathrm{Y}}}}+
  \alpha_{+}\tilde{k}^{2}\tilde{\mathcal{H}}_{{\rm B_{\mathrm{Y}}}},
  \nonumber
  \\
  \frac{\partial\tilde{\mathcal{H}}_{{\rm B_{\mathrm{Y}}}}}{\partial\tilde{\eta}} & =
  -2\tilde{k}^{2}\eta_{\mathrm{eff}}\tilde{\mathcal{H}}_{{\rm B_{\mathrm{Y}}}}+
  4\alpha_{-}\tilde{\mathcal{E}}_{{\rm B_{\mathrm{Y}}}},
  \nonumber
  \\
  \frac{\mathrm{d}\xi_{e\mathrm{R}}}{\mathrm{d}\tilde{\eta}} & =
  -\frac{3\alpha'}{\pi}\int_{\tilde{k}_{\mathrm{min}}}^{\tilde{k}_{\mathrm{max}}}\mathrm{d}\tilde{k}
  \frac{\partial\tilde{\mathcal{H}}_{{\rm B_{\mathrm{Y}}}}}{\partial\eta}-
  \Gamma(\xi_{e\mathrm{R}}-\xi_{e\mathrm{L}}+\xi_{0}),
  \nonumber
  \\
  \frac{\mathrm{d}\xi_{e\mathrm{L}}}{\mathrm{d}\tilde{\eta}} & =
  \frac{3\alpha'}{4\pi}\int_{k_{\mathrm{min}}}^{\tilde{k}_{\mathrm{max}}}\mathrm{d}\tilde{k}
  \frac{\partial\tilde{\mathcal{H}}_{{\rm B_{\mathrm{Y}}}}}{\partial\eta}-
  \frac{\Gamma}{2}(\xi_{e\mathrm{L}}-\xi_{e\mathrm{R}}-\xi_{0})-
  \frac{\Gamma_{\mathrm{sph}}}{2}\xi_{e\mathrm{L}},
  \nonumber
  \\
  \frac{\mathrm{d}\xi_{0}}{\mathrm{d}\tilde{\eta}} & =
  -\frac{\Gamma}{2}(\xi_{e\mathrm{R}}+\xi_{0}-\xi_{e\mathrm{L}}),
\end{align}
where $\tilde{\mathcal{E}}_{{\rm B_{\mathrm{Y}}}}(\tilde{k},\tilde{\eta})$
and $\tilde{\mathcal{H}}_{{\rm B_{\mathrm{Y}}}}(\tilde{k},\tilde{\eta})$
are the dimensionless spectral densities of the HMF energy and helicity.
They are related to $\rho_{\mathrm{Y}}^{(c)}(q,\xi)$ and $h_{\mathrm{Y}}^{(c)}(k,\xi)$
by $\rho_{\mathrm{Y}}^{(c)}(k,\xi)=\tilde{\mathcal{E}}_{{\rm B_{\mathrm{Y}}}}(\tilde{k},\tilde{\eta})T_{0}^{3}$
and $h_{\mathrm{Y}}^{(c)}(k,\xi)=\tilde{\mathcal{H}}_{{\rm B_{\mathrm{Y}}}}(\tilde{k},\tilde{\eta})T_{0}^{2}$,
where $T_{0}=2.7\,\text{K}$ is the temperature of the cosmic microwave
background radiation. The dimensionless conformal time $\tilde{\eta}$
and the conformal momentum $\tilde{k}$ in eq.~(\ref{eq:HMFsys}) are
$\eta=(2t_{\mathrm{Univ}}T_{0}/\tilde{M}_{\mathrm{Pl}})\tilde{\eta}$
and $k=T_{0}\tilde{k}$. Here $\tilde{M}_{\mathrm{Pl}}=M_{\mathrm{Pl}}/1.66\sqrt{g_{*}}$
is the effective Planck mass, $g_{*}=106.75$ is the number of the
relativistic degrees of freedom in the hot plasma before EWPT.

The asymmetries of right and left fermions $\xi_{e\mathrm{R,L}}$,
as well as that of Higgs bosons $\xi_{0}$, are $\xi_{e\mathrm{R,L}}=6(n_{e\mathrm{R,L}}-n_{\bar{e}\mathrm{R,L}})/T^{3}$and
$\xi_{0}=3(n_{\varphi_{0}}-n_{\bar{\varphi}_{0}})/T^{3}$, where $n_{(e,\bar{e})(\mathrm{R,L)}}$
are the number densities of right electrons, left fermions, their
antiparticles, $n_{\varphi_{0},\bar{\varphi}_{0}}$ are the number
densities of Higgs bosons and antibosons, and $T$ is the plasma temperature.
We account for only the lightest lepton generation since other
leptons are out of equilibrium sooner because of their greater Yukawa
coupling constants~\cite{Cam92}. The contribution of left fermions
is taken into account to make the system in eq.~(\ref{eq:HMFsys})
self consistent~\cite{DvoSem13}.

The spin-flip rate $\Gamma$ because of the interaction of fermions
with Higgs bosons reads~\cite{Cam92}
\begin{equation}\label{eq:spinfliprate}
  \Gamma(\eta)=\frac{242}{\tilde{\eta}_{\mathrm{EW}}}
  \left[
    1-\frac{\tilde{\eta}^{2}}{\tilde{\eta}_{\mathrm{EW}}^{2}}
  \right],
  \quad
  \tilde{\eta}_{\mathrm{EW}}=\frac{\tilde{M}_{\mathrm{Pl}}}{T_{\mathrm{EW}}}=7\times10^{15},
\end{equation}
where $T_{\mathrm{EW}}=10^{2}\,\text{GeV}$ is the temperature of
EWPT. The dimensionless sphaleron transitions rate is $\Gamma_{\mathrm{sph}}=8\times10^{-7}$~\cite{GorRub11}.

We choose the initial time of the HMFs evolution as $t_{*}=t_{\mathrm{RL}}$
which corresponds to $T=T_{\mathrm{RL}}=10\,\text{TeV}.$ Note that
both $\eta$ and $\tilde{\eta}$ are vanishing at $T=T_{\mathrm{RL}}$.
Below this temperature, Higgs decays become faster than universe expansion.
Thus, left fermions start to be produced. The minimal wave vector
$\tilde{k}_{\mathrm{min}}$, which corresponds to the maximal length
scale, is chosen as the inverse conformal horizon size at $T=T_{\mathrm{RL}}$,
$\tilde{k}_{\mathrm{min}}\sim T_{\mathrm{RL}}/\tilde{M}_{\mathrm{Pl}}\approx10^{-14}$.
The maximal wave vector $\tilde{k}_{\mathrm{max}}$, which is related
to the minimal length scale, is the free parameter in our model. The
minimal scale is chosen to be greater than the Debye length to satisfy
the plasma electroneutrality. We shall vary $\tilde{k}_{\mathrm{max}}$
in the range $10^{-5}<\tilde{k}_{\mathrm{max}}<10^{-2}$. In this
situation, the minimal length scale of HMFs is still greater than
the conformal Debye length $\tilde{r}_\mathrm{D}$.

The effective magnetic diffusion coefficient $\eta_{\mathrm{eff}}$ and the
effective $\alpha$-dynamo parameters $\alpha_{\pm}$ account for
the analogs of both the CME and the (H)MHD turbulence for HMFs. They
are~\cite{DvoSem17}
\begin{align}\label{eq:etaalpha}
  \eta_{\mathrm{eff}} & =
  \sigma_{c}^{-1}+\frac{4}{3}\frac{(\alpha')^{-2}}{\tilde{\rho}+\tilde{p}}
  \int_{\tilde{k}_{\mathrm{min}}}^{\tilde{k}_{\mathrm{max}}}\mathrm{d}\tilde{k}\mathcal{\tilde{E}}_{{\rm B_{\mathrm{Y}}}},
  \quad\alpha'=\frac{g'^{2}}{4\pi},
  \nonumber
  \\
  \alpha_{\pm} & =
  \alpha_{\mathrm{Y}}(\tilde{\eta})\mp\frac{2}{3}\frac{(\alpha')^{-2}}{\tilde{\rho}+\tilde{p}}
  \int_{\tilde{k}_{\mathrm{min}}}^{\tilde{k}_{\mathrm{max}}}\mathrm{d}\tilde{k}\tilde{k}^{2}\mathcal{\tilde{H}}_{{\rm B_{\mathrm{Y}}}},
\end{align}
where $\sigma_{c}\approx10^{2}$ is the conformal conductivity of
relativistic plasma, $\tilde{\rho}$ and $\tilde{p}$ are the plasma
density and pressure expressed in conformal variables, $g'=e/\cos\theta_{\mathrm{W}}$
is the hypercharge, and
\begin{equation}\label{eq:alpha}
  \alpha_{\mathrm{Y}}(\tilde{\eta})=\frac{\alpha'}{\pi\sigma_{c}}
  \left[
    \xi_{e\mathrm{R}}(\tilde{\eta})-\frac{\xi_{e\mathrm{L}}(\tilde{\eta})}{2}
  \right],
\end{equation}
is the $\alpha$-dynamo parameter when only the analog of the CME
for HMFs is accounted for. We choose $p=\rho/3$ for the ultrarelativistic
plasma. The value of $\alpha'=9.5\times10^{-3}.$ Note that $\alpha_{\mathrm{Y}}$ in eq.~\eqref{eq:alpha} is analogous to the $\alpha$-parameter, responsible for the instability of Maxwell magnetic fields in the early universe after EWPT owing to the CME, which is used in ref.~\cite{Rog17}.

HMF can be unstable if $\alpha_\mathrm{Y}$ is nonzero. It is the consequence of the analog of the CME in which the (hyper-)electric current is induced along the (hyper-)magnetic field. Besides the CME, one has the chiral vortical effect (the CVE) in the system of ultrarelativistic fermions. The CVE is the generation of a current $\mathbf{j}_\mathrm{CVE} = \tfrac{\mu_5\mu}{\pi^2}\bm{\omega}$~\cite{Kha16} along the plasma vorticity $\bm{\omega} = \tfrac{1}{2}(\nabla\times\mathbf{v})$. Taking into account that $\mathbf{v} \propto (\mathbf{j}\times\mathbf{B})$ in the chosen model for (H)MHD turbulence~\cite{Sig02}, as well as $(\nabla\times\mathbf{B}) = \mathbf{j} + \mathbf{j}_\mathrm{CVE}$, one gets that the CVE contribution to the induction equation is quadratic in HMF. Thus, such a contribution is cubic in HMF in the equations for the HMF energy density and the helicity, which are the binary combinations of HMFs. Supposing that random HMFs are Gaussian, i.e. all odd correlators are vanishing, one gets that the CVE does not contribute to eq.~\eqref{eq:HMFsys}.

The initial condition for eq.~(\ref{eq:HMFsys}) is chosen as $\xi_{e\mathrm{L}}=\xi_{0}=0$
and $\xi_{e\mathrm{R}}=10^{-10}$. Such values give the appropriate
value of the observed $\text{BAU} \sim10^{-10}$ if this model
is applied for this problem~\cite{DvoSem21}. The initial spectra
are $\mathcal{\tilde{E}}_{{\rm B_{\mathrm{Y}}}}^{(0)}(\tilde{k})=\mathcal{C}_{\mathrm{Y}}\tilde{k}^{n_{\mathrm{Y}}}$
and $\mathcal{\tilde{H}}_{{\rm B_{\mathrm{Y}}}}^{(0)}(\tilde{k})=2q\mathcal{\tilde{E}}_{{\rm B_{\mathrm{Y}}}}^{(0)}(\tilde{k})/\tilde{k}$,
where $0<q<1$ is the phenomenological parameter fixing the helicity
of a seed HMF. The constant $\mathcal{C}_{\mathrm{Y}}$ is chosen
such that 
\begin{equation}
  \tilde{B}_{\mathrm{Y}}^{(0)}=
  \left[
    2\int_{\tilde{k}_{\mathrm{min}}}^{\tilde{k}_{\mathrm{max}}}\mathrm{d}\tilde{k}
    \mathcal{\tilde{E}}_{{\rm B_{\mathrm{Y}}}}^{(0)}(\tilde{k})
  \right]^{1/2}
\end{equation}
is the dimensionless strength of a seed HMF. We vary $\tilde{B}_{\mathrm{Y}}^{(0)}$
in the range $1.4\times10^{-2}<\tilde{B}_{\mathrm{Y}}^{(0)}<1.4\times10^{-1}$.
Such HMFs contribute to neither the universe expansion nor the primordial
nucleosynthesis~\cite{DvoSem21}.

We take the initial Kolmogorov spectrum with $n_{\mathrm{Y}}=-5/3$. The Kolmogorov
spectrum has the singularity at $\tilde{k} = 0$ which has to be regularized by setting the lower bound $\tilde{k} > \tilde{k}_\mathrm{min}$, which corresponds to the maximal length scale, $k_\mathrm{min} \propto L_\mathrm{max}^{-1}$. As we mentioned above, we choose $L_\mathrm{max}$ comparable with the horizon size at $T = T_\mathrm{RL}$ (see also ref.~\cite{DvoSem21}). A more realistic choice of the seed spectrum was used in ref.~\cite{Rop20}. It consists in the Batchelor spectrum $\mathcal{\tilde{E}}_{{\rm B_{\mathrm{Y}}}}^{(0)}(\tilde{k}) \propto \tilde{k}^4$ for small $\tilde{k}$ below a certain value and the Kolmogorov one above it.

The energy of plasma dissipates into heat because of viscous processes if the typical length scale is less than the Kolmogorov one $L_\mathrm{visc} \propto \nu^{3/4}$~\cite{Dav15}, where $\nu$ is the kinematic viscosity coefficient. In our approach, we keep only the Lorentz force in the Navier-Stokes equation and neglect other terms including the viscous one~\cite{Sig02}. Therefore we can take that $L_\mathrm{visc}$ is vanishing. It means that the only relevant lower bound for the minimal length scale is determined by the Debye length, $\tilde{k}_\mathrm{max}^{-1} = \tilde{L}_\mathrm{min} \gg \tilde{r}_\mathrm{D}$.

For the derivation of eqs.~(\ref{eq:HMFsys})-(\ref{eq:alpha}) and
the discussion of their applicability for the description of the HMFs
evolution the reader is referred to refs.~\cite{DvoSem13,DvoSem21,DvoSem17}.
Note that the coefficients analogous to that in eq.~(\ref{eq:etaalpha})
were also studied in ref.~\cite{Cam07}. We can also rewrite eqs.~(\ref{eq:rhoGWisotr})-(\ref{eq:HMFsys})
in the new variables convenient for numerical simulations. The details
are provided in appendix~\ref{sec:NEWVAR}.

\section{Results\label{sec:RES}}

In this section, we show the behavior of the energy density of GWs
based on the numerical solution of eq.~(\ref{eq:newsys}) with the
parameters and the initial condition adopted in section~\ref{sec:HMFEVOL}.

\begin{figure}
  \centering
  \subfigure[]
  {\label{fig:diffBa}
  \includegraphics[scale=.33]{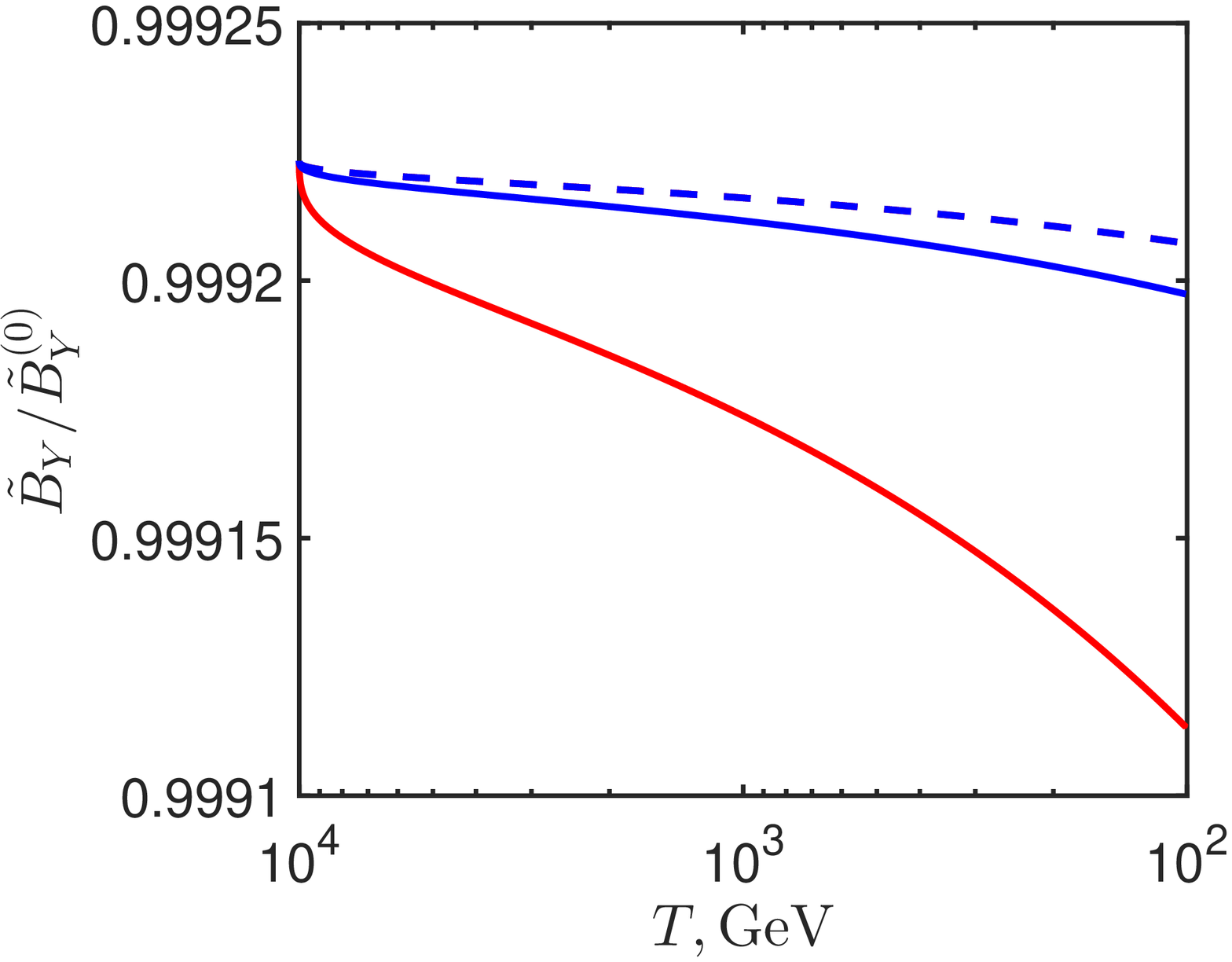}}
  \hskip-.6cm
  \subfigure[]
  {\label{fig:diffBb}
  \includegraphics[scale=.33]{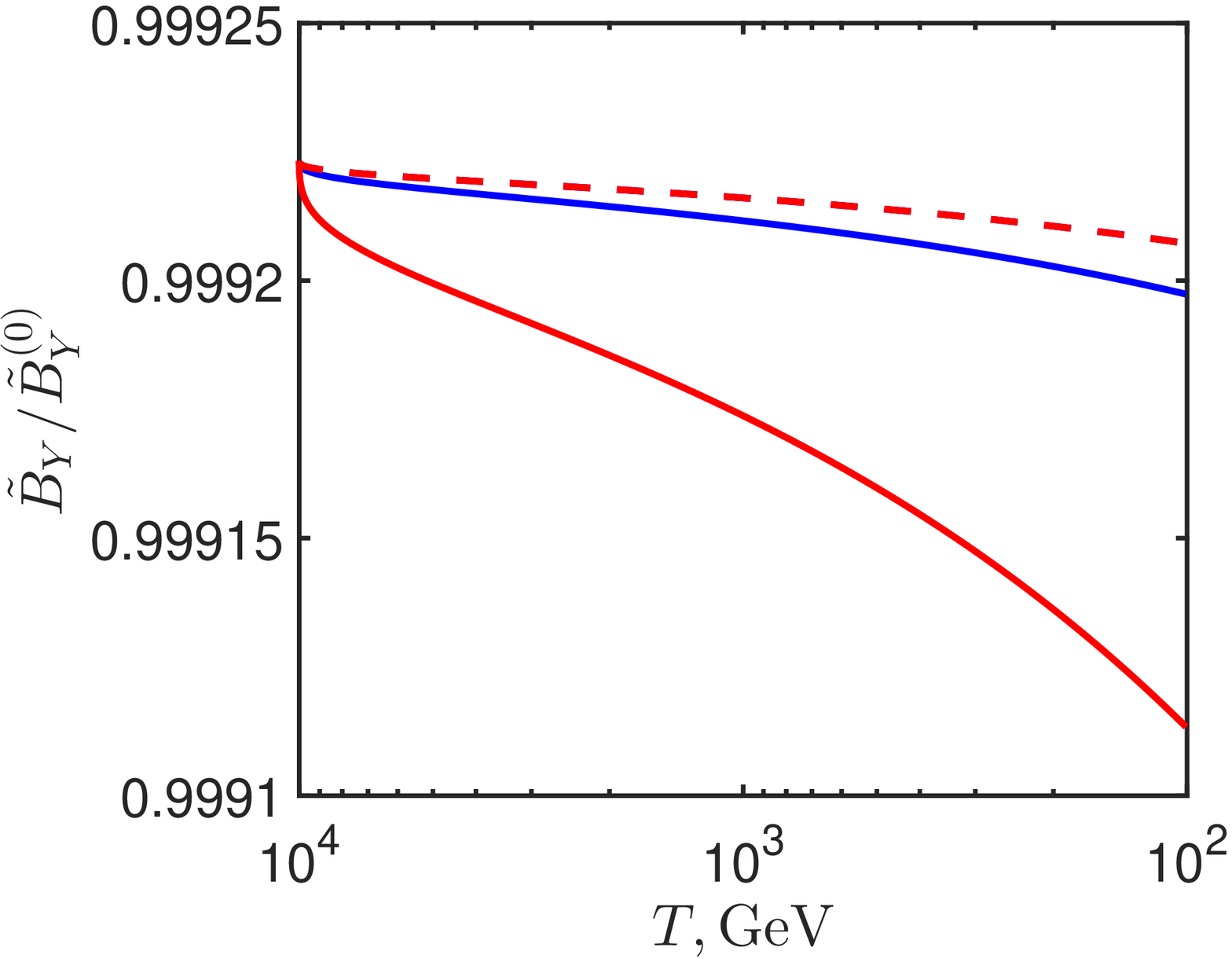}}
  \\
  \subfigure[]
  {\label{fig:diffBc}
  \includegraphics[scale=.33]{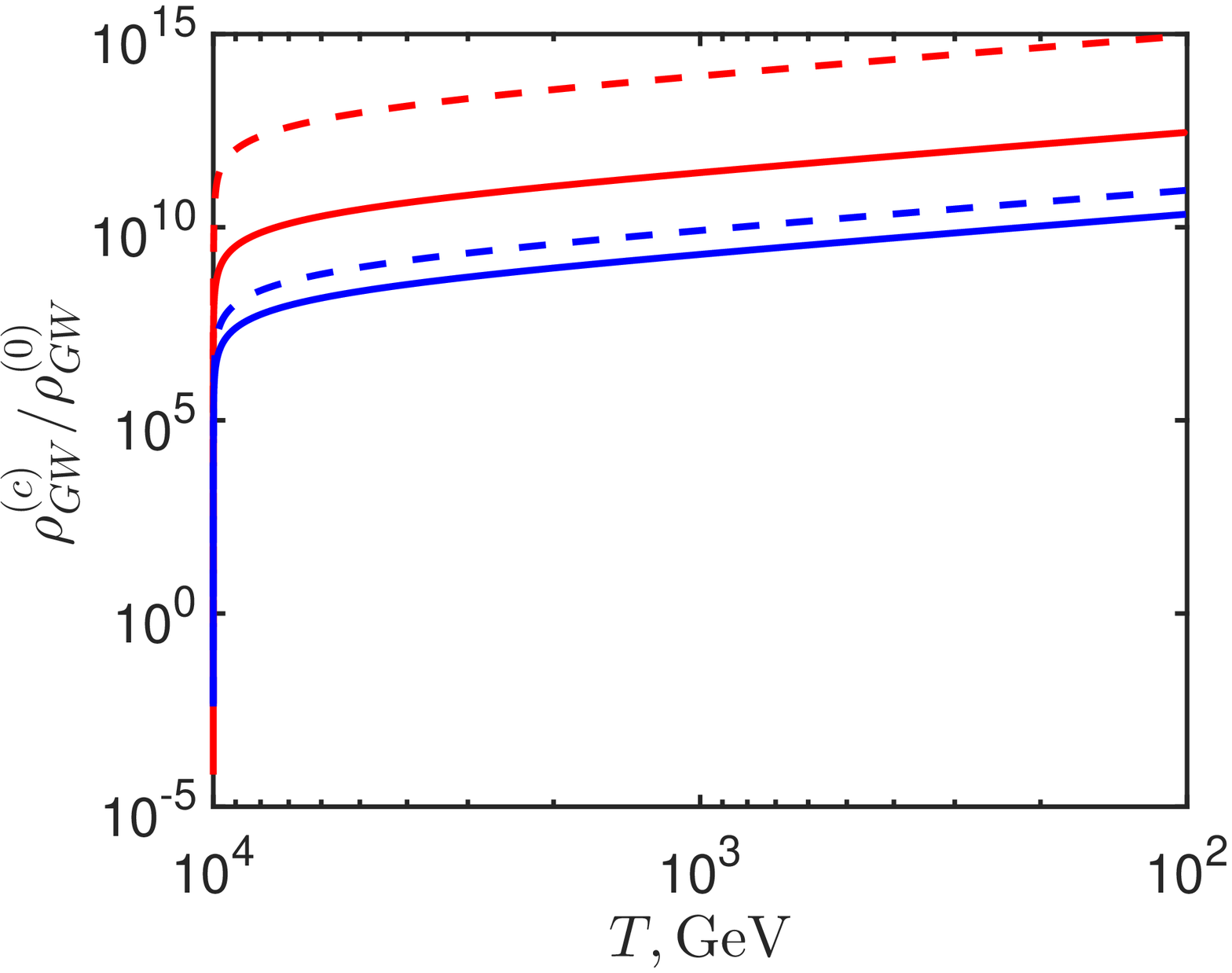}}
  \hskip-.6cm
  \subfigure[]
  {\label{fig:diffBd}
  \includegraphics[scale=.33]{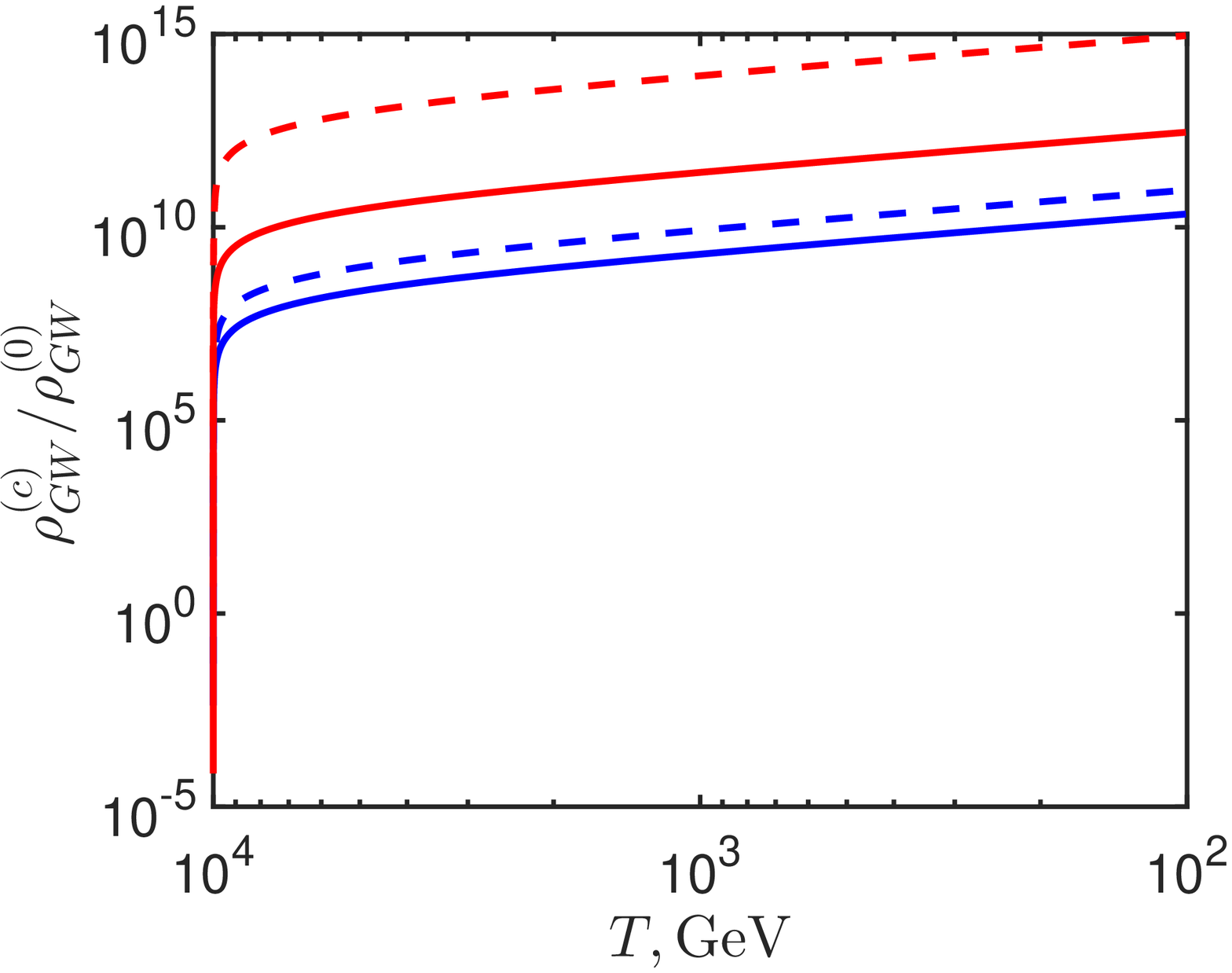}}
  \\
  \subfigure[]
  {\label{fig:diffBe}
  \includegraphics[scale=.33]{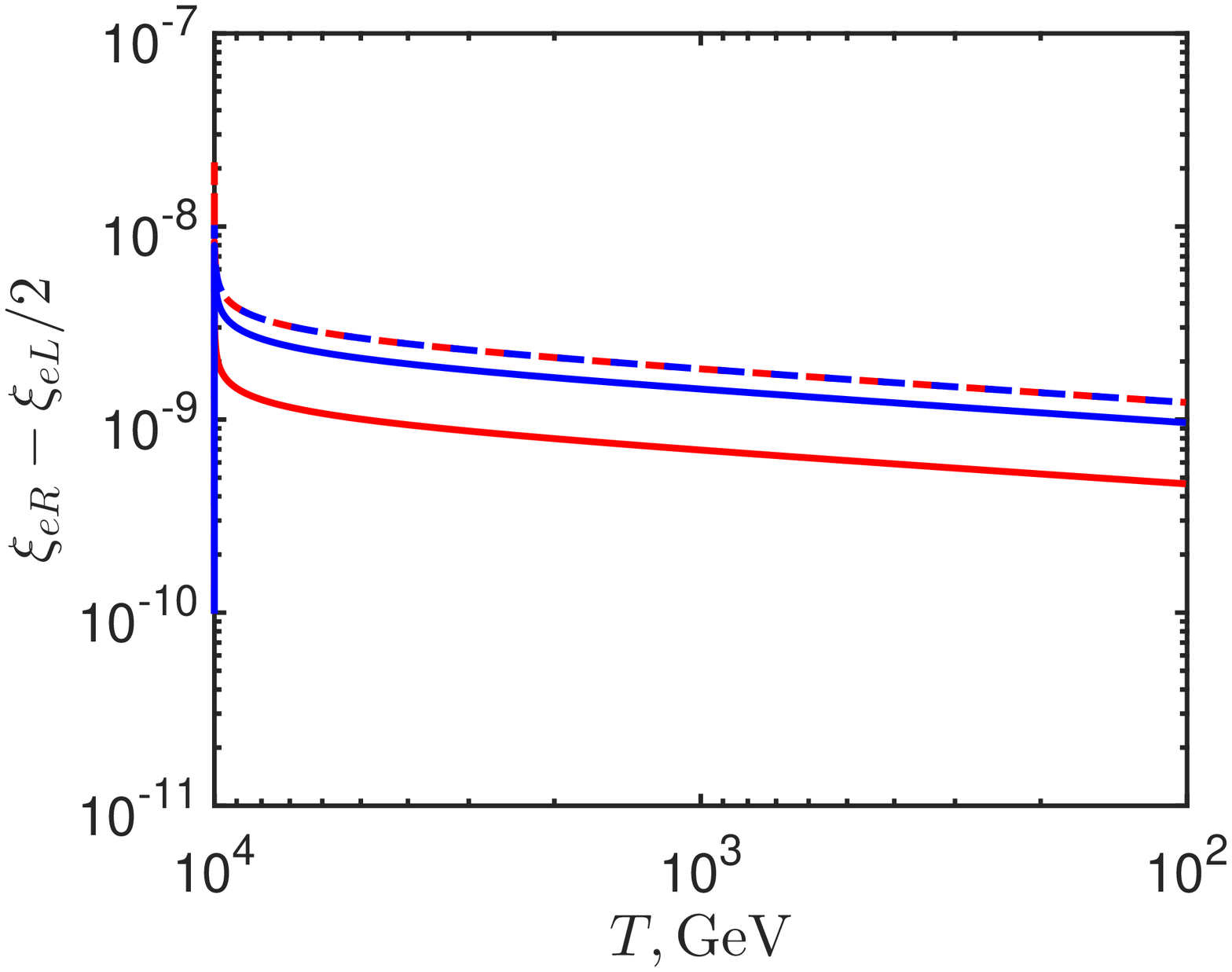}}
  \hskip-.6cm
  \subfigure[]
  {\label{fig:diffBf}
  \includegraphics[scale=.33]{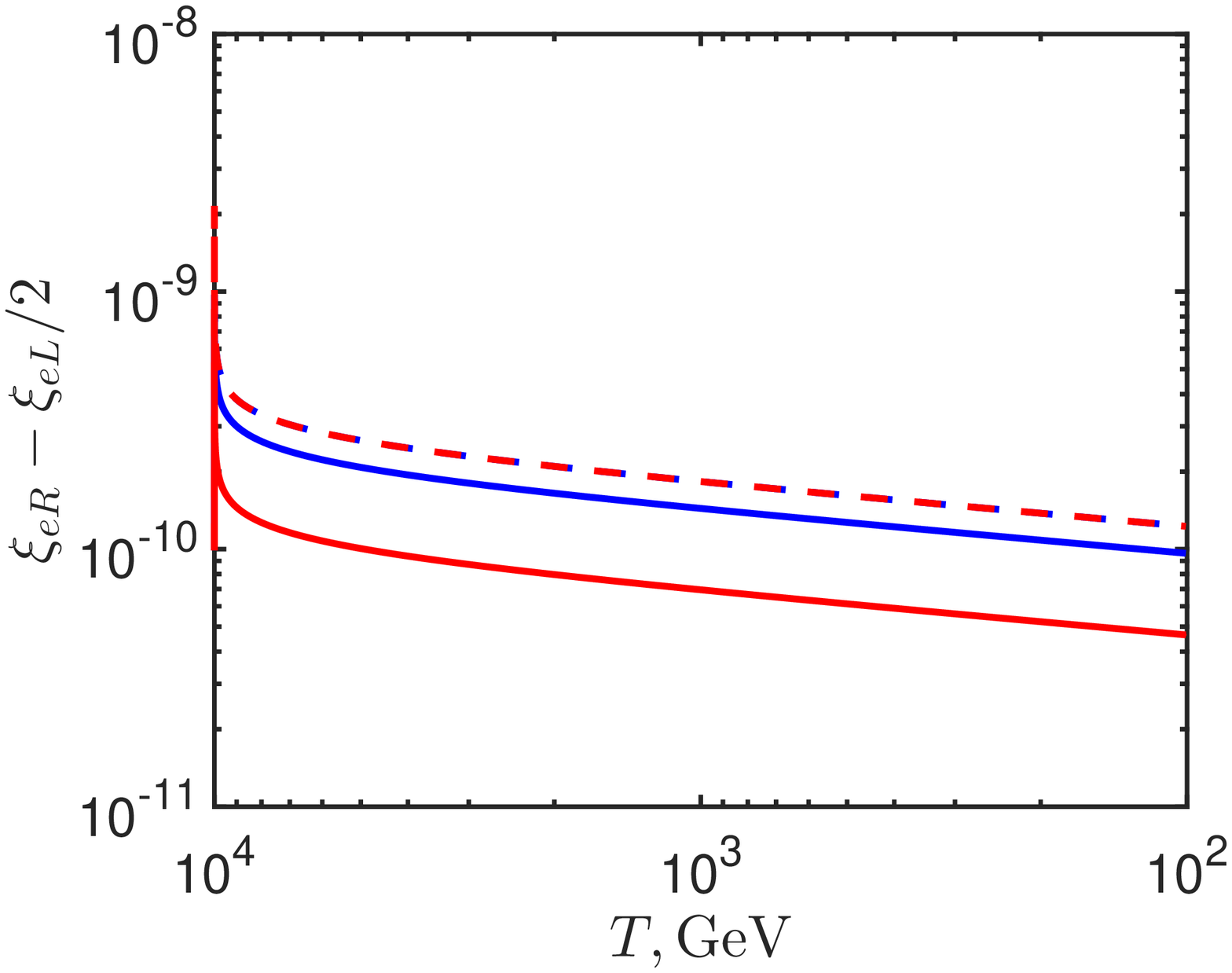}}
  \protect
\caption{The evolution of the system at different $\tilde{B}_{\mathrm{Y}}^{(0)}$
and $q$ for the fixed $\tilde{k}_{\mathrm{max}}=10^{-3}$ based on
the numerical solution of eq.~(\ref{eq:newsys}). The initial condition
is $\xi_{e\mathrm{R}}^{(0)}=10^{-10}$, $\xi_{e\mathrm{L}}^{(0)}=\xi_{0}^{(0)}=0$,
and $\rho_{\mathrm{GW}}^{(c)}=0$ at $T_{\mathrm{RL}}=10\,\text{TeV}$.
We account for the (H)MHD turbulence in solid lines, whereas the dashed
ones are plotted without taking the turbulence into account. Red lines
correspond to $\tilde{B}_{\mathrm{Y}}^{(0)}=1.4\times10^{-1}$ and
blue lines to $\tilde{B}_{\mathrm{Y}}^{(0)}=1.4\times10^{-2}$. Panels~(a)
and~(b): the evolution of the HMF; panels~(c) and~(d): the evolution
of the energy density of GWs; cf. eqs.~(\ref{eq:densGW}) and~(\ref{eq:rho0});
panels~(e) and~(f): the evolution of the $\alpha$-dynamo parameter
$\sim\xi_{e\mathrm{R}}-\xi_{e\mathrm{L}}/2$. In panels (a), (c),
and~(e), we set $q=1$; and panels (b), (d), and~(f) correspond
to $q=0.1$.\label{fig:diffB}}
\end{figure}

In figure~\ref{fig:diffB}, we present the evolution of the system
from $T=T_{\mathrm{RL}}$ down to EWPT for the fixed $\tilde{k}_{\mathrm{max}}=10^{-3}$
and different $\tilde{B}_{\mathrm{Y}}^{(0)}$ and $q$. Comparing
figures~\ref{fig:diffBa} and~\ref{fig:diffBb}, where the plots
of HMF are shown, we can see that there is a very small dependence
of the results on~$q$.

The HMF evolution depends on whether we account for the (H)MHD turbulence
or not. The HMF evolution, without taking the (H)MHD turbulence into
account, corresponds to $\eta_{\mathrm{eff}}=\sigma_{c}^{-1}$ and
$\alpha_{\pm}=\alpha_{\mathrm{Y}}$ in eq.~(\ref{eq:etaalpha}).
The main contribution of the HMFs noise was mentioned in ref.~\cite{DvoSem21}
to result from $\eta_{\mathrm{eff}}$. Thus, the greater $\tilde{B}_{\mathrm{Y}}^{(0)}$
is, the faster HMF decays. This feature can be seen in figures~\ref{fig:diffBa}
and~\ref{fig:diffBb}. Red and blue dashed lines, plotted without
the (H)MHD turbulence, overlap. It should be noted that, despite red
solid lines are below blue ones, the absolute value of HMFs, shown
in red, is almost one order of magnitude greater than of these depicted
in blue in figures~\ref{fig:diffBa} and~\ref{fig:diffBb}.

The evolution of the energy density of GWs $\rho_{\mathrm{GW}}^{(c)}$
in the cooling universe is shown in figures~\ref{fig:diffBc} and~\ref{fig:diffBd}.
First, greater $\tilde{B}_{\mathrm{Y}}^{(0)}$ leads to greater $\rho_{\mathrm{GW}}^{(c)}$;
cf. red and blue solid lines in figures~\ref{fig:diffBc} and~\ref{fig:diffBd}.
Second, greater $\tilde{B}_{\mathrm{Y}}^{(0)}$ result in the faster
decay of HMFs. Thus the energy density of GWs, which depends on $\tilde{B}_{\mathrm{Y}}$,
will evolve slower for greater $\tilde{B}_{\mathrm{Y}}^{(0)}$. This
feature can be also observed figures~\ref{fig:diffBc} and~\ref{fig:diffBd}:
the gap between solid and dashed lines in red is wider than between
these in blue. Third, $\rho_{\mathrm{GW}}^{(c)}$ grows while the
universe cools down and HMFs decays; cf. figures~\ref{fig:diffBa}
and~\ref{fig:diffBb}. It is a consequence of the cumulative effect
of the integration over the conformal time in eq.~(\ref{eq:rhoGWisotr}).

The behavior of the $\alpha$-dynamo parameter $\alpha_{\mathrm{Y}}\sim\xi_{e\mathrm{R}}-\xi_{e\mathrm{L}}/2$
is depicted in figures~\ref{fig:diffBe} and~\ref{fig:diffBf}.
In fact, $\alpha_{\mathrm{Y}}$ is almost completely determined by
the evolution of $\xi_{e\mathrm{R}}$ since $|\xi_{e\mathrm{L}}|\ll\xi_{e\mathrm{R}}$,
as found in ref.~\cite{DvoSem21}.

One can see in figures~\ref{fig:diffBe} and~\ref{fig:diffBf} that $\alpha_\mathrm{Y}$ has a spike at $T\gtrsim T_\mathrm{RL}$. It can be explained by the backreaction of the hypermagnetic helicity on the asymmetries. Indeed, this spike results from the term $I_\mathrm{H}$ in eq.~\eqref{eq:newsys} for the asymmetry of right electrons $M_\mathrm{R}$. The level of the spike in higher for the greater initial helicity. This feature is seen in figures~\ref{fig:diffBe} and~\ref{fig:diffBf}. The asymmetries fall down at greater evolution times because of the Higgs decays and especially the sphaleron process which acts on the asymmetry of left fermions. Taking into account the Higgs decays and the sphaleron process results in the nonconservation of the sum of the chiral imbalance and the (hyper-)magnetic helicity, which is usually conserved if only the Adler anomaly is considered. These processes are analogous to the spin-flip of fermions in electromagnetic plasma~\cite{Dvo16,Boy21}. The evolution of HMF  in figures~\ref{fig:diffBa} and~\ref{fig:diffBb} does not reveal the spikes at $T\gtrsim T_\mathrm{RL}$ since it is dominated by the diffusion terms $\propto -2\tilde{k}^{2} \sigma_c^{-1} \tilde{\mathcal{E}}_{{\rm B_{\mathrm{Y}}}}$ in eq.~\eqref{eq:HMFsys}.

The obtained behavior of HMFs can be explained if we calculate the typical scale of the HMFs instability. Using the results of ref.~\cite{Semikoz:2016lqv}, one gets that
\begin{equation}\label{eq:kappastar}
  \tilde{k}_\star = \frac{\pi \sigma_c}{2\alpha'}
  \left(
    \int_0^{\tilde{\eta}} \mathrm{d}\tilde{\eta} \xi_{e\mathrm{R}}
  \right)^{-1},  
\end{equation}
where we keep only the right asymmetry for simplicity. Equation~\eqref{eq:kappastar} can be obtained, e.g., from the induction equation for HMF, $\partial_\eta\tilde{\mathbf{B}}_\mathrm{Y} = \alpha_\mathrm{Y} (\nabla\times \tilde{\mathbf{B}}_\mathrm{Y}) + \dotsb$, where we keep only the instability term. Assuming the Chern-Simons wave distribution and using eq.~\eqref{eq:alpha}, one gets that $\tilde{B}_\mathrm{Y} \propto \exp(\tilde{k}/\tilde{k}_\star)$. A more careful derivation of $\tilde{k}_\star$ is provided in ref.~\cite{Semikoz:2016lqv}.

One expects the HMF amplification if $\tilde{k} > \tilde{k}_\star$. The fast growth of $\xi_{e\mathrm{R}}$ to the value $\sim 10^{-8}$ takes place in the temperature range $\Delta T = T_\mathrm{RL} - T = 10^{-1}\,\text{GeV}$ (see, e.g., figure~\ref{fig:diffBe} and ref.~\cite{DvoSem21}). Thus we can estimate $\tilde{k}_\star \sim 10^3$, which is much greater than $\tilde{k}_\mathrm{max} = 10^{-3}$ used in figure~\ref{fig:diffB}. It means that the diffusion is dominant in the HMF evolution. To trigger a sizable instability of HMFs in the considered system, one has to consider, e.g., a greater value of $\tilde{k}_\mathrm{max} \sim \tilde{k}_\star$. However, the condition of the plasma electroneutrality is violated even for $\tilde{k}_\mathrm{max} = 0.1 \sim \tilde{r}_\mathrm{D}^{-1}$.

\begin{figure}
  \centering
  \subfigure[]
  {\label{fig:speca}
  \includegraphics[scale=.33]{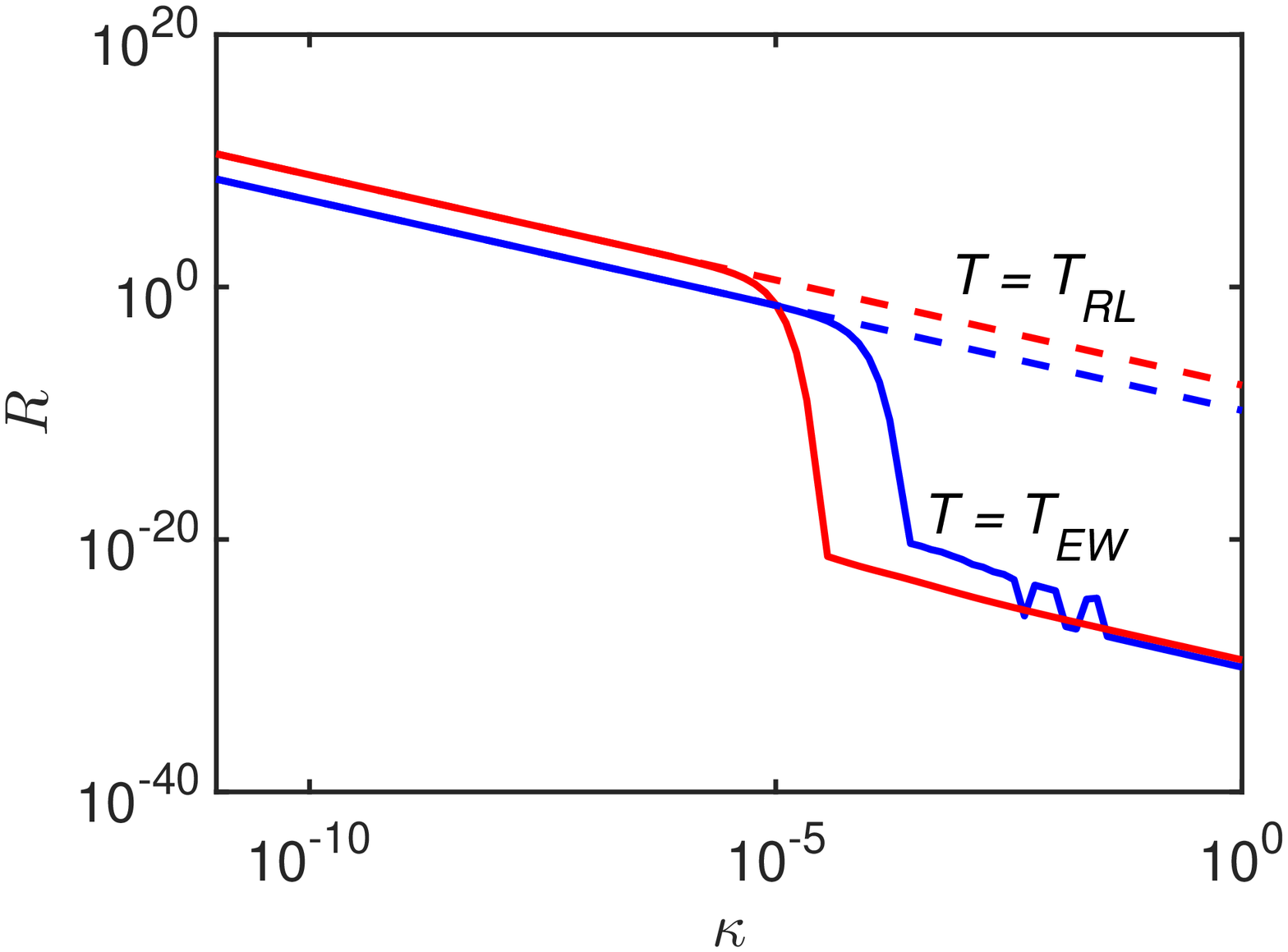}}
  \hskip-.6cm
  \subfigure[]
  {\label{fig:specb}
  \includegraphics[scale=.33]{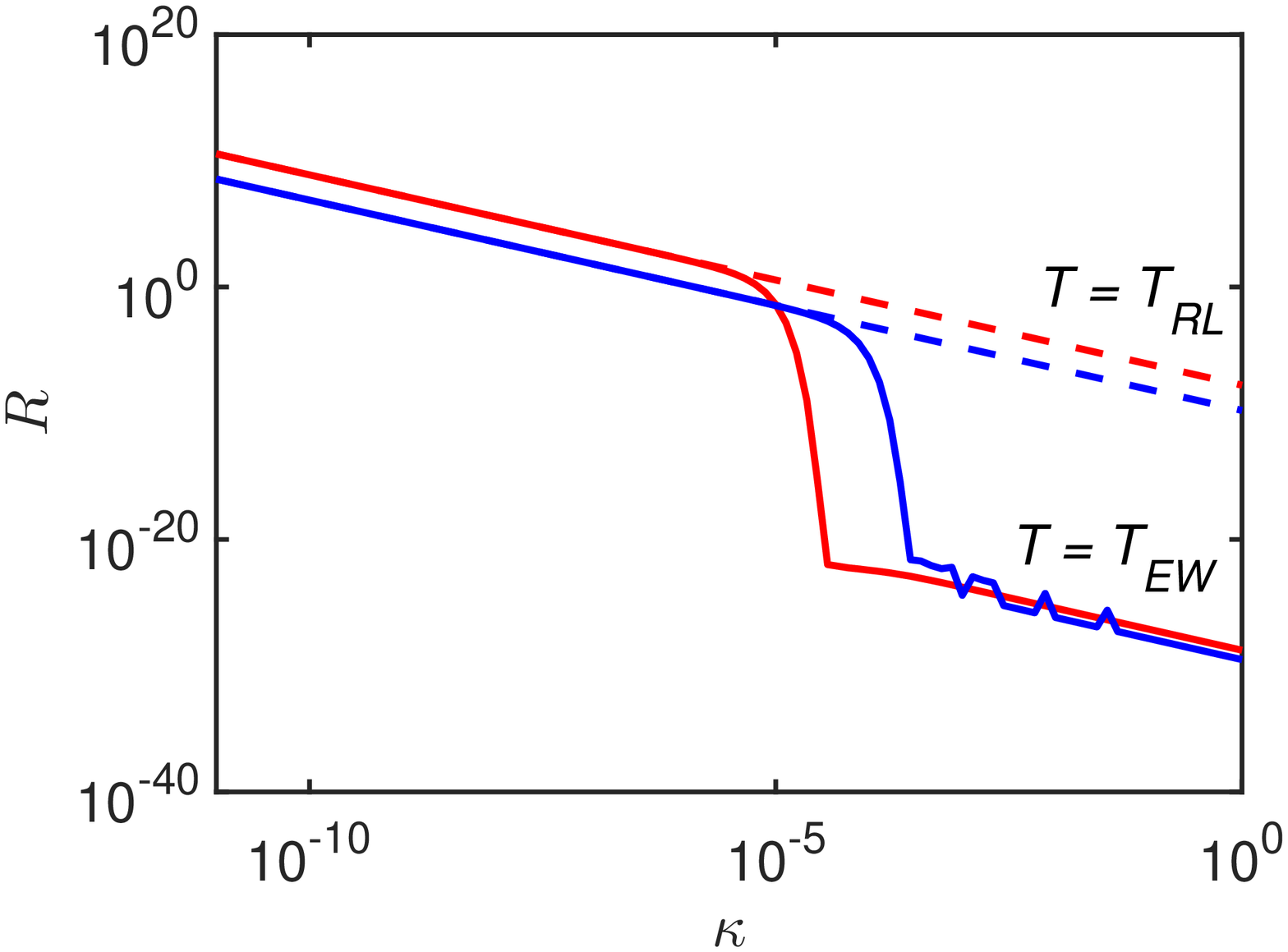}}
  \\
  \subfigure[]
  {\label{fig:specc}
  \includegraphics[scale=.33]{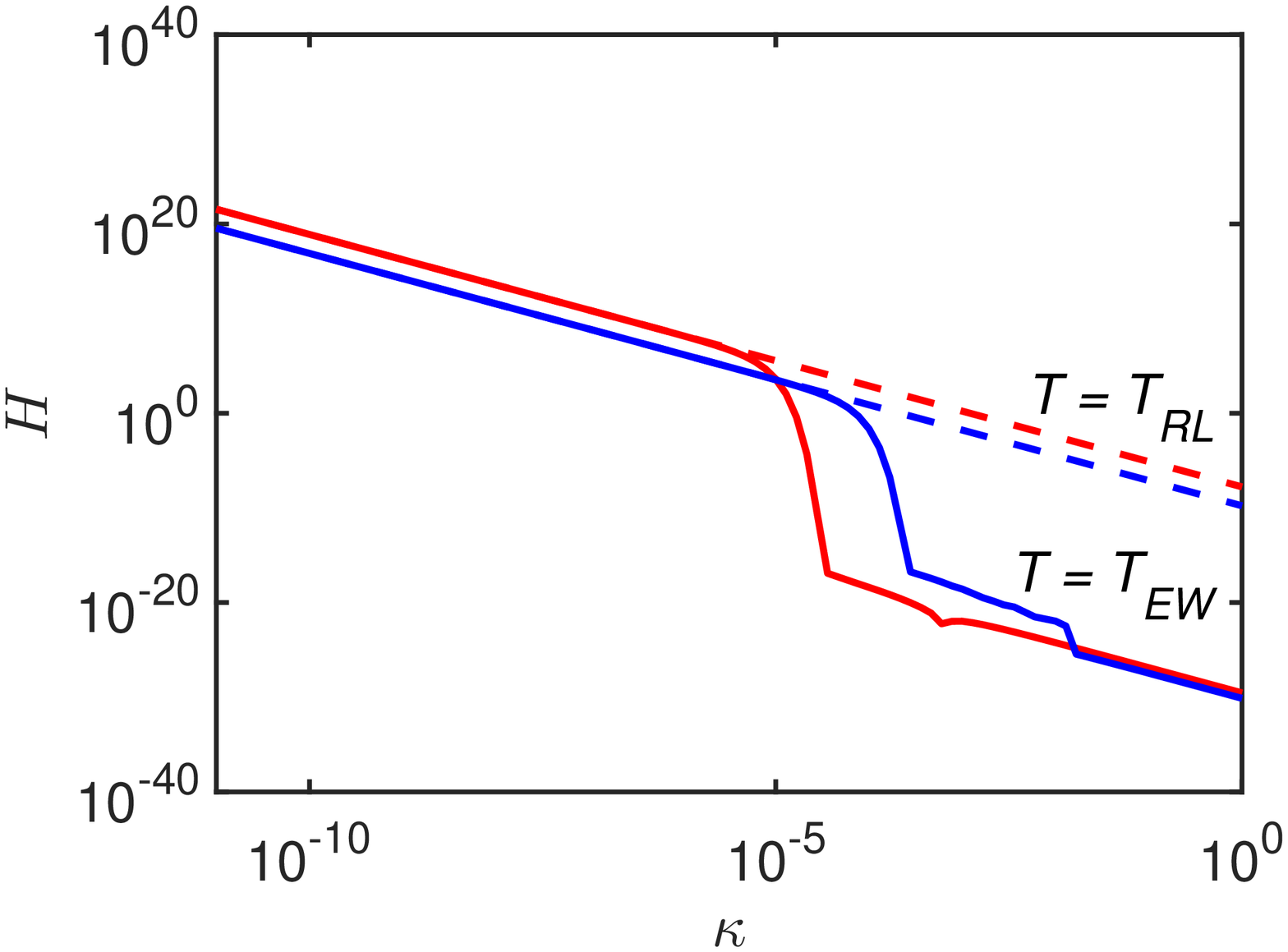}}
  \hskip-.6cm
  \subfigure[]
  {\label{fig:specd}
  \includegraphics[scale=.33]{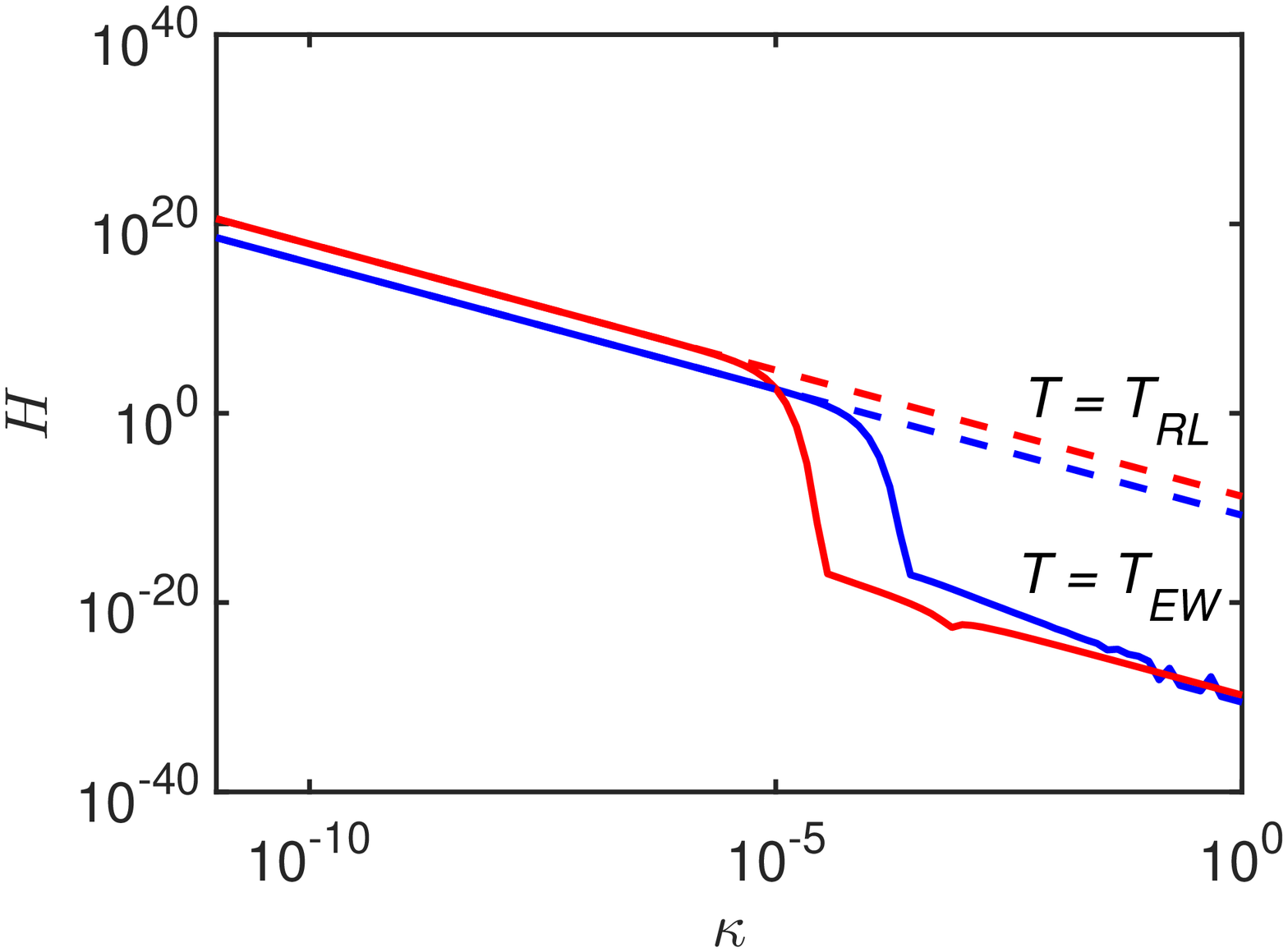}}
  \protect
\caption{The spectra of the hypermagnetic energy $R$ and the hypermagnetic helicity $H$ versus the dimensioless wave vector $\kappa$ (see eq.~\eqref{eq:newvar}) for different $\tilde{B}_{\mathrm{Y}}^{(0)}$
and $q$, as well as the fixed $\tilde{k}_{\mathrm{max}}=10^{-3}$. The initial condition
is the same as in figure~\ref{fig:diffB}. Here, we account for the (H)MHD turbulence.
Dashed lines show the spectra of seed HMFs at $T=T_\mathrm{RL}$. Solid lines represent the situation when the universe cools down to EWPT.
Red lines correspond to $\tilde{B}_{\mathrm{Y}}^{(0)}=1.4\times10^{-1}$ and
blue lines to $\tilde{B}_{\mathrm{Y}}^{(0)}=1.4\times10^{-2}$. Panels~(a)
and~(b): the spectra of the energy density; panels~(c) and~(d): the spectra of the hypermagnetic helicity;
In panels (a) and~(c), we set $q=1$; and panels (b) and~(d) correspond
to $q=0.1$.\label{fig:spec}}
\end{figure}

The spectra of the hypermagnetic energy $R$ and the hypermagnetic helicity $H$ (see eq.~\eqref{eq:newvar}) at different plasma temperatures are depicted in figure~\ref{fig:spec}. We show the initial spectra at $T=T_\mathrm{RL}$, which are Kolmogorov and depicted with dashed lines, and the spectra at EWPT. The spectra of seed fields evolve to the solid lines very rapidly. Therefore we do not show them for any intermediate temperatures. 

The spectra at $T = T_\mathrm{EW}$ coincide with the seed ones, shown with dashed lines, for $\kappa \lesssim 10^{-5}$. For greater $\kappa$, the spectra start to decrease. This behavior is accounted for by the diffusion terms $\propto -\kappa^2$ in eq.~\eqref{eq:newsys}. Thus, modes with greater $\kappa$ are damped more efficiently. This evolution of the HMFs is promising from the observational point of view since HMFs with large scales survive. Moreover, we can see in figure~\ref{fig:spec} that the spectra corresponding to a stronger seed field, shown with red color, start to decrease at smaller $\kappa$ compared to blue ones. It happens owing to the turbulence contribution to the effective diffusion coefficient. The above features justify our suggestion that the evolution of HMFs in this model is dominated by the diffusion.

The spectra at $T = T_\mathrm{EW}$ demonstrate the drastic change in their behavior when they drop below $\sim 10^{-20}$. This feature is related to the limited accuracy of numerical simulations. Thus, we can rely only on the part of the curves in figure~\ref{fig:spec} which are above $\sim 10^{-20}$.
 
\begin{figure}
  \centering
  \subfigure[]
  {\label{fig:diffka}
  \includegraphics[scale=.33]{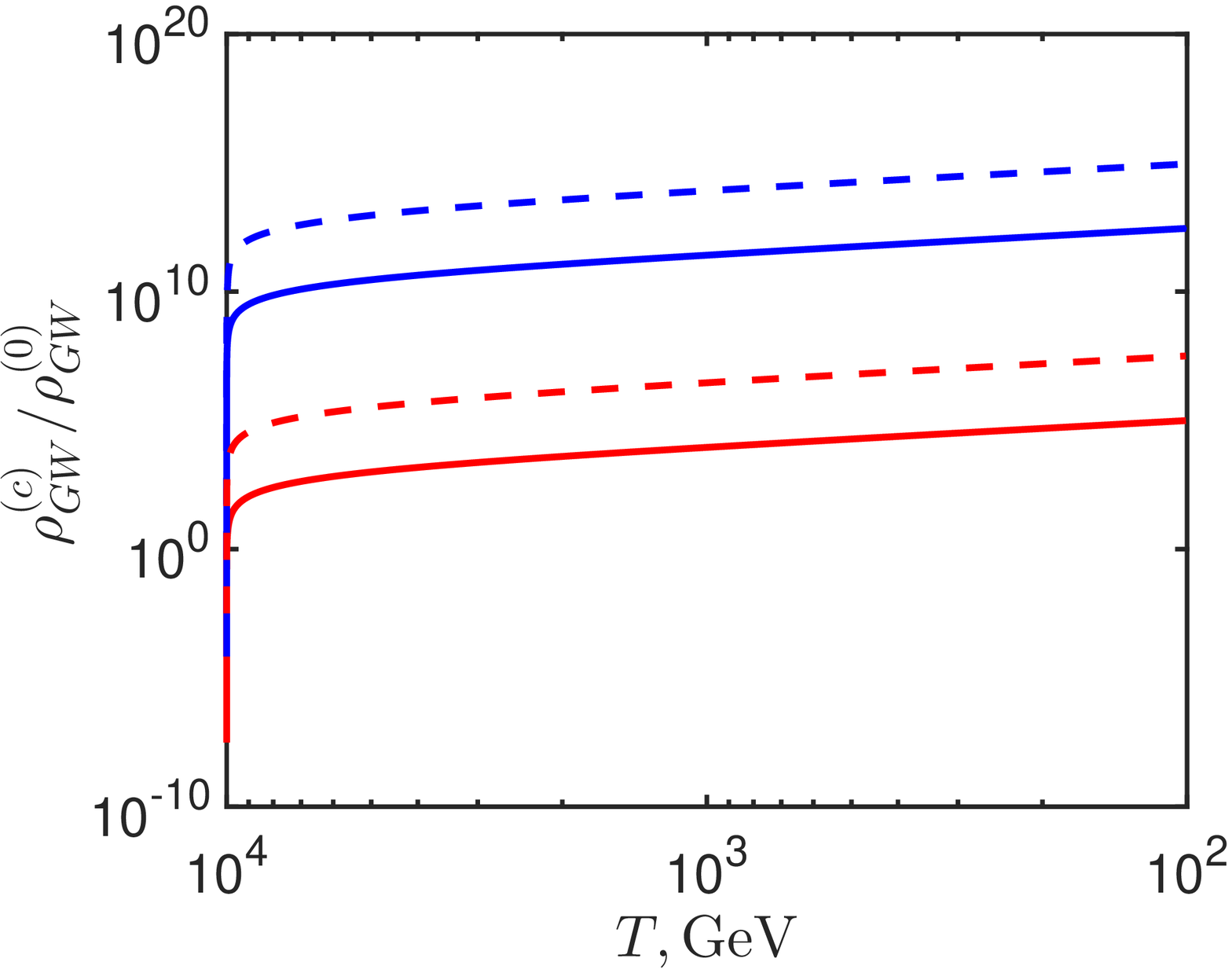}}
  \hskip-.6cm
  \subfigure[]
  {\label{fig:diffkb}
  \includegraphics[scale=.33]{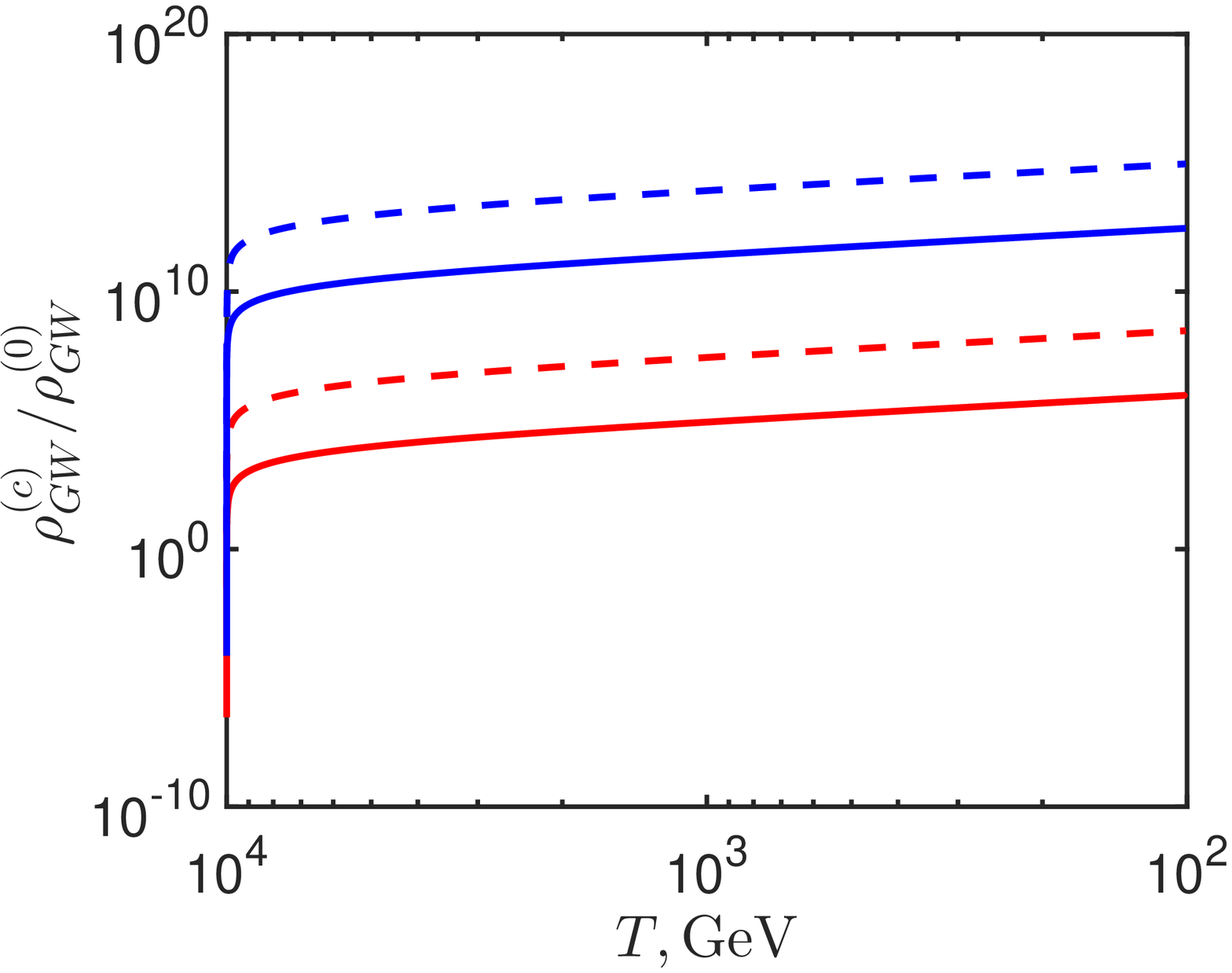}}
  \protect
\caption{The evolution of the energy density of GWs at different $\tilde{k}_{\mathrm{max}}$
and $q$ for the fixed $\tilde{B}_{\mathrm{Y}}^{(0)}=1.4\times10^{-1}$
based on the numerical solution of eq.~(\ref{eq:newsys}). The initial
condition is the same as in figure~\ref{fig:diffB}. We account for
the (H)MHD turbulence in solid lines, wheres the dashed ones are plotted
without taking the turbulence into account. Red lines correspond to
$\tilde{k}_{\mathrm{max}}=10^{-2}$ and blue lines to $\tilde{k}_{\mathrm{max}}=10^{-3}$.
(a) $q=1$; and (b) $q=0.1$.\label{fig:diffk}}
\end{figure}

The influence of the HMFs scale on the production of GWs is shown
in figure~\ref{fig:diffk}, where we fix the strength of the seed HMF
by $\tilde{B}_{\mathrm{Y}}^{(0)}=1.4\times10^{-1}$ and vary $\tilde{k}_{\mathrm{max}}$.
The effect of the (H)MHD turbulence on the evolution of HMFs was mentioned
in refs.~\cite{DvoSem21,DvoSem17} to be more sizable for small scale
fields, i.e. for great $\tilde{k}_{\mathrm{max}}$. Thus HMFs, corresponding
to greater $\tilde{k}_{\mathrm{max}}$ decay faster. It results in
the slower enhancement of $\rho_{\mathrm{GW}}^{(c)}$ for such fields.
Such a behavior is seen in figure~\ref{fig:diffk} where red lines,
with $\tilde{k}_{\mathrm{max}}=10^{-2}$, are below blue ones, with
$\tilde{k}_{\mathrm{max}}=10^{-3}$. We can also see the negligible
dependence on the initial hypermagnetic helicity; cf. figures~\ref{fig:diffka}
and~\ref{fig:diffkb}.

The evolution of the system, shown in figures~\ref{fig:diffB} and~\ref{fig:diffk},
qualitatively resembles the results obtained in ref.~\cite{Bra21},
where the generation of primordial GWs is driven by the CME. However,
unlike ref.~\cite{Bra21}, where the turbulence was modeled fully
numerically, we use a semi-analytical approach, which allows one to
analyze the influence of different factors, like the strength of a
seed HMF and its scale, on the production of GWs. Moreover, we study
more realistic situation when the generation of relic GWs is driven
by the lepton and Higgs asymmetries before EWPT. For this purpose, we utilize
the analogs of the CME for the HMFs and of the Adler anomalies for
the asymmetries.

\subsection{Observability of relic GWs\label{subsec:OBS}}

\begin{figure}
\centering
\includegraphics[scale=0.33]{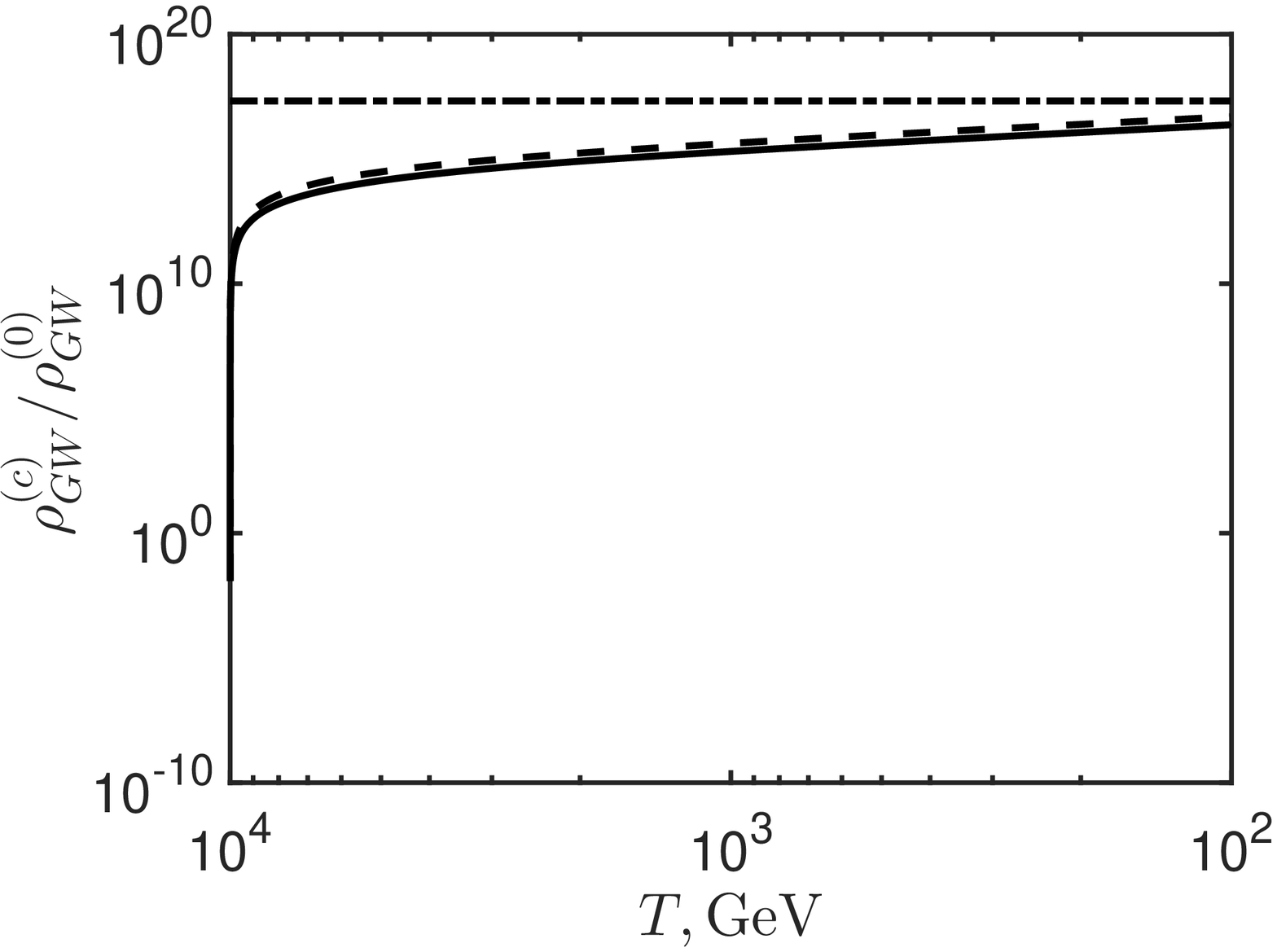}
\protect
\caption{The energy density of GWs versus the plasma temperature for $\tilde{B}_{\mathrm{Y}}^{(0)}=1.4\times10^{-1}$
($B_{\mathrm{Y}}^{(0)}=7\times10^{26}\,\text{G}$), $\tilde{k}_{\mathrm{max}}=10^{-5}$,
and $q=1$. In the solid line, we account for the (H)MHD turbulence,
whereas, in dashed one, not. The rest of the parameters and the initial
condition are the same as in figure~\ref{fig:diffB}. The horizontal dash-dotted line is the averaged lower bound for the GW energy density, which can be experimentally probed presently.\label{fig:obs}}
\end{figure}

Let us consider the possibility for the predicted GW signal to be
observed with the current experimental techniques. For this purpose
we take that $\tilde{B}_{\mathrm{Y}}^{(0)}=1.4\times10^{-1}$, $\tilde{k}_{\mathrm{max}}=10^{-5}$,
and $q=1$. The chosen $B_{\mathrm{Y}}^{(0)}=7\times10^{26}\,\text{G}$
still does not violate the Big Bang nucleosynthesis constraint $B_{\mathrm{BBN}}=10^{11}\text{G}$
at $T_\mathrm{BBN}=0.1\,\text{MeV}$~\cite{CheSchTru94}. The value of $\tilde{k}_{\mathrm{max}}$
is taken so that the HMFs noise is still noticeable but its influence
is small (see solid and dashed lines in figure~\ref{fig:obs}). The
rest of the parameters is the same as in figure~\ref{fig:diffB}.

One can see in the end of the solid line at $T=10^{2}\,\text{GeV}$
in figure~\ref{fig:obs} that the conformal energy density of GWs at
$T_{\mathrm{EW}}$ is $\rho_{\mathrm{GW}}^{(c)}(T_{\mathrm{EW}})=5\times10^{15}\rho_{\mathrm{GW}}^{(0)}=7.5\times10^{-8}\,\text{eV}\cdot\text{cm}^{-3}$,
where $\rho_{\mathrm{GW}}^{(0)}$ is given in eq.~(\ref{eq:rho0}).
We also assume that no other sources of GWs are after EWPT. Thus the
current energy density of GWs is $\rho_{\mathrm{GW}}^{(\mathrm{now})}=\rho_{\mathrm{GW}}^{(c)}/a_{\mathrm{now}}^{3}=\rho_{\mathrm{GW}}^{(c)}(T_{\mathrm{EW}})$.

Some modern GWs detectors (see, e.g., ref.~\cite{Arz20}) can potentially
detect stochastic GWs with $\Omega\sim10^{-10}$, where $\Omega=\tfrac{f\rho_{\mathrm{GW}}(f)}{\rho_{\mathrm{crit}}}$,
$f=k/2\pi$ is the frequency of GWs in Hz, $\rho_{\mathrm{GW}}(f)$ is the spectrum of the energy density with respect to $f$, and $\rho_{\mathrm{crit}}=0.5\times10^{-5}\,\text{GeV}\cdot\text{cm}^{-3}$
is the critical energy density of the universe. We can evaluate the total observable energy density as $\rho_{\mathrm{GW}}^{(\mathrm{obs})}\sim3.1\times10^{-6}\,\text{eV}\cdot\text{cm}^{-3}$,
which is only $40$ times greater than the predicted value $\rho_{\mathrm{GW}}^{(\mathrm{now})}=7.5\times10^{-8}\,\text{eV}\cdot\text{cm}^{-3}$.
Thus, the GW background, described within our model, is potentially detectable
using current GWs detectors after some enhancement of their sensitivity.

We can see in figure~\ref{fig:obs} that both the turbulent GW energy density, shown by the solid line, and the nonturbulent one, depicted by the dashed line, are below the observations threshold, represented by the dash-dotted line. The value of $(\rho_\mathrm{GW}^{(c)}/\rho_\mathrm{GW}^{(0)})_\mathrm{obs} = 2.1\times10^{17}$ corresponds to the sensitivity of, e.g., the NANOGrav experiment~\cite{Arz20}. It corresponds to $\Omega\sim 10^{-10}.$

\section{Conclusion\label{sec:CONCL}}

In this work, we have studied the production of relic GWs in turbulent
plasma in the symmetric phase before the EWPT. Random HMFs, with the
(H)MHD turbulence, are the driver for the GWs generation. HMFs are
taken to couple to the gravity through their energy-momentum tensor.
The evolution of HMFs in primordial plasma is governed by the analogs
of the CME and the Adler anomalies in the presence of nonzero asymmetries
of right and left electrons, left neutrinos, their antiparticles,
as well as Higgs bosons.

We have used the realistic parameters and the initial condition of
HMFs for the production of GWs. The same parameters, while utilized in
the problem of the BAU generation (see, e.g., ref.~\cite{DvoSem21}),
result in BAU close to the observed value, $(\text{BAU})_{\mathrm{obs}}\sim10^{-10}$.
It is the advantage of our work compared to ref.~\cite{Bra21}, where
the analogous problem, i.e. the generation of relic GWs driven by
the CME, was studied. Moreover, we use the semi-analytical model,
which is based on the (H)MHD turbulence~\cite{Sig02,Cam07,DvoSem17}.
It allows one to guess the dependence of the final results on the
parameters of the system, in contrast to ref.~\cite{Bra21}, where
purely numerical simulations were used.

In section~\ref{sec:PRODGW}, we have rederived eq.~(\ref{eq:Dprime})
for the tensor perturbations of the metric in the presence of HMFs.
Then, using its formal solution in eq.~(\ref{eq:Dsol}), we have
obtained the expression for the spectrum of the energy density of
GWs in eq.~(\ref{eq:rhockern}) represented in conformal variables.
After averaging and using the binary combinations of HMFs, such as
their energy and the helicity, we get the final expressions for the
energy density of GWs and its spectrum in eqs.~(\ref{eq:densGW})
and~(\ref{eq:rhoGWisotr}).

Section~\ref{sec:HMFEVOL} is devoted to the formulation of the dynamics
of HMFs and the initial condition for them. Basically, it is similar
to that in ref.~\cite{DvoSem21}, where we studied the BAU generation
driven by turbulent HMFs. Such HMFs evolve owing to the analog of
the CME in the presence of nonzero lepton asymmetries. We have also
accounted for the backreaction of helical HMFs to the lepton asymmetries
evolution because of the analog of the Adler anomalies for HMFs. The
noise of HMFs is modeled by the analog of the (H)MHD turbulence. The
main kinetic equations for the binary combinations of HMFs and the
particle asymmetries are summarized in eqs.~(\ref{eq:HMFsys})-(\ref{eq:alpha}).

In section~\ref{sec:RES}, we have presented the results of the numerical
solution of the evolution eq.~(\ref{eq:HMFsys}) and the energy density
of GWs in eq.~(\ref{eq:densGW}) basing on this solution. In figures~\ref{fig:diffB}
and~\ref{fig:diffk}, we have shown the strength of HMF $\tilde{B}_{\mathrm{Y}}$,
the energy density of GWs $\rho_{\mathrm{GW}}^{(c)}$, and the $\alpha$-dynamo
parameter $\sim\xi_{e\mathrm{R}}-\xi_{e\mathrm{L}}/2$ in the temperature
range from $T=T_{\mathrm{RL}}=10\,\text{TeV}$ down to $T=T_{\mathrm{EW}}=10^{2}\,\text{GeV}$.
We have analyzed the dependence of $\rho_{\mathrm{GW}}^{(c)}$ on
the strength of the seed HMF $\tilde{B}_{\mathrm{Y}}^{(0)}$ for the
fixed minimal scale $\tilde{k^{-1}}_{\mathrm{max}}$ (see figure~\ref{fig:diffB}),
as well as on the minimal scale for the fixed strength of the seed
HMF (see figure~\ref{fig:diffk}). There is a negligible dependence
of $\rho_{\mathrm{GW}}^{(c)}$ on the initial helicity of HMFs. There
results qualitatively resemble the findings of ref.~\cite{Bra21}.

We have briefly discussed the possibility to observe the predicted
GWs background in section~\ref{subsec:OBS}. Using the following realistic
initial conditions: $B_{\mathrm{Y}}^{(0)}=7\times10^{26}\,\text{G}$,
$\tilde{k}_{\mathrm{max}}=10^{-5}$, $q=1$, $\xi_{e\mathrm{R}}^{(0)}=10^{-10}$,
and $\xi_{e\mathrm{L}}^{(0)}=\xi_{0}^{(0)}=0$, we have obtained that
the current energy density of GWs produced is only $\sim40$ times
below the threshold of the modern GWs detectors. Thus such a signal
can be potentially observable in the nearest future.

In appendix~\ref{sec:NEWVAR}, we have rewritten the main expressions
in the form adapted for numerical simulations.

\acknowledgments

%\section*{Acknowledgements}
I am thankful to A.~Yu.~Smirnov and V.~B.~Semikoz for useful discussions.

\appendix

\section{New variables\label{sec:NEWVAR}}

Following ref.~\cite{DvoSem21}, it is
convenient to use the new variables in eq.~(\ref{eq:HMFsys}),
\begin{align}\label{eq:newvar}
  \tilde{\mathcal{E}}_{{\rm B_{\mathrm{Y}}}}(\tilde{k},\tilde{\eta}) & =
  \frac{\tilde{k}_{\mathrm{max}}\pi^{2}}{6\alpha'^{2}}R(\kappa,\tau),
  \quad
  \tilde{\mathcal{H}}_{{\rm B_{\mathrm{Y}}}}(\tilde{k},\tilde{\eta})=\frac{\pi^{2}}{3\alpha'^{2}}H(\kappa,\tau),
  \nonumber
  \\
  \xi_{\mathrm{R,L,0}}(\tilde{\eta}) & =
  \frac{\pi\tilde{k}_{\mathrm{max}}}{\alpha'}M_{\mathrm{R,L,0}}(\tau),
  \quad
  \tau=\frac{2\tilde{k}_{\mathrm{max}}^{2}}{\sigma_{c}}\tilde{\eta},
  \quad
  \tilde{k}=\tilde{k}_{\mathrm{max}}\kappa,
\end{align}
 where $\kappa_{m}<\kappa<1$, $\kappa_{m}=\tilde{k}_{\mathrm{min}}/\tilde{k}_{\mathrm{max}}$,
and $\tau\geq0$ is the new dimensionless time. Using eq.~(\ref{eq:newvar}),
we rewrite eq.~(\ref{eq:HMFsys}) in the form~\cite{DvoSem21},
\begin{align}\label{eq:newsys}
  \frac{\partial R}{\partial\tau}= & -\kappa^{2}
  \left(
    1+\lambda_t I_\mathrm{R}
  \right)R +
  \kappa^{2}
  \left(
    M_{\mathrm{R}}-\frac{M_{\mathrm{L}}}{2}-\lambda_t I_\mathrm{H}
  \right)H,
  \nonumber
  \displaybreak[2]
  \\
  \frac{\partial H}{\partial\tau}= & -\kappa^{2}
  \left(
    1+\lambda_t I_\mathrm{R}
  \right)H +
  \left(
    M_{\mathrm{R}}-\frac{M_{\mathrm{L}}}{2}+\lambda_t I_\mathrm{H}
  \right)R,
  \nonumber
  \displaybreak[2]
  \\
  \frac{\mathrm{d}M_{\mathrm{R}}}{\mathrm{d}\tau}= &
  I_\mathrm{H}-
  \left(
    M_{\mathrm{R}}-\frac{M_{\mathrm{L}}}{2}
  \right)
  I_\mathrm{R} -
  \Gamma'(M_{\mathrm{R}}-M_{\mathrm{L}}+M_{0}),
  \nonumber
  \displaybreak[2]
  \\
  \frac{\mathrm{d}M_{\mathrm{L}}}{\mathrm{d}\tau}= & 
  -\frac{1}{4}I_\mathrm{H}+\frac{1}{4}
  \left(
    M_{\mathrm{R}}-\frac{M_{\mathrm{L}}}{2}
  \right)
  I_\mathrm{R} -
  \Gamma'(M_{\mathrm{L}}-M_{\mathrm{R}}-M_{0})/2-\frac{\Gamma'_{s}}{2}M_{\mathrm{L}},
  \nonumber
  \displaybreak[2]
  \\
  \frac{\mathrm{d}M_{0}}{\mathrm{d}\tau}= & -\Gamma'(M_{\mathrm{R}}+M_{0}-M_{\mathrm{L}})/2,
\end{align}
where 
\begin{align}\label{eq:chiral}
  I_\mathrm{R}(\tau) = & \int_{\kappa_{m}}^{1}\mathrm{d}\kappa'R(\kappa',\tau),
  \quad
  I_\mathrm{H}(\tau) = \int_{\kappa_{m}}^{1}\mathrm{d}\kappa'\kappa'^{2}H(\kappa',\tau),
  \quad
  \lambda_t = \frac{2\sigma_{c}\tilde{k}_{\mathrm{max}}^{2}\pi^{2}}{9\alpha'^{4}(\tilde{p}+\tilde{\rho})},
  \notag
  \\
  \Gamma'(\tau)=&\frac{121\sigma_{c}}{\tilde{\eta}_{\mathrm{EW}}\tilde{k}_{\mathrm{max}}^{2}}
  \left[
    1-\frac{T_{\mathrm{EW}}^{2}}{T_{\mathrm{RL}}^{2}}
    \left(
      1+\frac{T_{\mathrm{RL}}}{M_{0}}\frac{\sigma_{c}}{2\tilde{k}_{\mathrm{max}}^{2}}\tau
    \right)^{2}
  \right],
\end{align}
and $\Gamma'_{s}=\sigma_{c}\Gamma_{\mathrm{sph}}/2\tilde{k}_{\mathrm{max}}^{2}$.

Equation~(\ref{eq:newsys}) should be completed with the initial
condition, which has the form,
\begin{equation}\label{energy}
  R(\kappa,0)=C_{\mathrm{Y}}\kappa^{n_{\mathrm{Y}}},
  \quad
  H(\kappa,0)=q\frac{R(\kappa,0)}{\kappa},
\end{equation}
where
\begin{equation}\label{eq:CY}
  C_{\mathrm{Y}}=\frac{3\alpha'^{2}(1+n_{\mathrm{Y}})[\tilde{B}_{\mathrm{Y}}^{(0)}]^{2}}
  {\pi^{2}\tilde{k}_{\mathrm{max}}^{2}(1-\kappa_{m}^{1+n_{\mathrm{Y}}})}.
\end{equation}
Equations~(\ref{eq:newsys})-(\ref{eq:CY}) can be solved numerically.

The spectral density of GWs and their energy density in eqs.~(\ref{eq:rhoGWisotr})
and~(\ref{eq:densGW}) should be also adapted to the new variables.
The spectrum of the energy density is
\begin{align}\label{eq:specGWdmless}
  \rho_{\mathrm{GW}}^{(c)}(\kappa,\tau)= &
  \frac{\sigma_{c}^{2}\pi^{2}t_{\text{Univ}}^{2}T_{0}^{7}T_{\mathrm{RL}}^{2}G}
  {576\alpha'^{4}\tilde{M}_{\mathrm{Pl}}^{2}\tilde{k}_{\mathrm{max}}}
  \frac{\tau}{\kappa^{3}}
  \int_{0}^{\tau}\frac{\mathrm{d}\tau'}{(1+7.1\times10^{-13}\tau'/\tilde{k}_{\mathrm{max}})^{2}}
  \nonumber
  \\
  & \times
  \iint_{S(\kappa)}\frac{\mathrm{d}\upsilon\mathrm{d}\varpi}{\upsilon{}^{3}\varpi^{3}}
  \nonumber
  \\
  & \times
  \big\{
    [4\kappa^{2}\upsilon^{2}+(\kappa^{2}+\upsilon^{2}-\varpi^{2})^{2}]
    [4\kappa^{2}\varpi^{2}+(\kappa^{2}-\upsilon^{2}+\varpi^{2})^{2}]R(\upsilon,\tau')R(\varpi,\tau')
    \nonumber
    \\
    & +
    16\kappa^{2}\upsilon^{2}\varpi^{2}(\kappa^{2}+\upsilon^{2}-\varpi^{2})(\kappa^{2}-\upsilon^{2}+\varpi^{2})
    H(\upsilon,\tau')H(\varpi,\tau')
  \big\},
\end{align}
where $0<\kappa<2$ and $S(\kappa)$ is the 2D integration domain in the $(\upsilon,\varpi)$-plane.
The energy density reads
\begin{align}\label{eq:endensdmless}
  \rho_{\mathrm{GW}}^{(c)}(\tau)= & 
  \int_{0}^{2k_{\mathrm{max}}}\rho_{\mathrm{GW}}^{(c)}(k,\tau)\mathrm{d}k=
  \rho_{\mathrm{GW}}^{(0)}\tau\int_{0}^{\tau}\frac{\mathrm{d}\tau'}{(1+7.1\times10^{-13}\tau'/\tilde{k}_{\mathrm{max}})^{2}}
  \nonumber
  \\
  & \times
  \int_{0}^{2}\frac{\mathrm{d}\kappa}{\kappa^{3}}\iint_{S(\kappa)}\frac{\mathrm{d}\upsilon\mathrm{d}\varpi}{\upsilon{}^{3}\varpi^{3}} 
  \nonumber
  \\
  & \times
  \big\{
    [4\kappa^{2}\upsilon^{2}+(\kappa^{2}+\upsilon^{2}-\varpi^{2})^{2}]
    [4\kappa^{2}\varpi^{2}+(\kappa^{2}-\upsilon^{2}+\varpi^{2})^{2}]R(\upsilon,\tau')R(\varpi,\tau')
    \nonumber
    \\
    & +
    16\kappa^{2}\upsilon^{2}\varpi^{2}(\kappa^{2}+\upsilon^{2}-\varpi^{2})(\kappa^{2}-\upsilon^{2}+\varpi^{2})
    H(\upsilon,\tau')H(\varpi,\tau')
  \big\},
\end{align}
where 
\begin{equation}\label{eq:rho0}
  \rho_{\mathrm{GW}}^{(0)}=\frac{\sigma_{c}^{2}\pi^{2}t_{\text{Univ}}^{2}T_{0}^{8}T_{\mathrm{RL}}^{2}G}
  {576\alpha'^{4}\tilde{M}_{\mathrm{Pl}}^{2}}=
  1.5\times10^{-23}\,\text{eV}\cdot\text{cm}^{-3}.
\end{equation}
Equation~(\ref{eq:endensdmless}) is used to plot figures~\ref{fig:diffBc},
\ref{fig:diffBd}, \ref{fig:diffk}, and~\ref{fig:obs} basing
on the numerical solution, $R(\kappa,\tau)$ and $H(\kappa,\tau)$,
of eq.~(\ref{eq:newsys}).

\end{document}